\documentclass[lettersize,journal]{IEEEtran}
\usepackage{amsmath,amsfonts}
\usepackage{algorithmic}
\usepackage{algorithm}
\usepackage{array}
\usepackage[caption=false,font=normalsize,labelfont=sf,textfont=sf]{subfig}
\usepackage{textcomp}
\usepackage{stfloats}
\usepackage{url}
\usepackage{verbatim}
\usepackage{graphicx}
\usepackage{cite}
\usepackage{amssymb}
\usepackage{enumitem}

\usepackage{mathtools}

\usepackage{pdfpages}

\usepackage{tikz,pgf}
\usetikzlibrary{patterns}

\newtheorem{theorem}{Theorem}
\newtheorem{definition}{Definition}

\newtheorem{lem}{Lemma}
\newtheorem{claim}{Claim}
\newtheorem{example}{Example}
\newtheorem{remark}{Remark}

\usepackage{bbm}
\DeclareMathOperator{\Tr}{Tr}

\begin{document}

\title{Successive Approximation Coding for Distributed Matrix Multiplication}

\author{Shahrzad~Kiani,~\IEEEmembership{Graduate~Student~Member,~IEEE,}
        and~Stark~C.~Draper,~\IEEEmembership{Senior~Member,~IEEE}
        % <-this % stops a space
\thanks{This work was supported in part by a Discovery Research Grant from the Natural Sciences and Engineering Research Council of Canada (NSERC) and by a NSERC Alexander Graham Bell Canada Graduate Scholarship-Doctoral
(CGS D3).}
\thanks{S. Kiani and S. C. Draper are with the Department of Electrical and Computer  Engineering, University of Toronto, Toronto, ON, Canada (Emails: shahrzad.kianidehkordi@mail.utoronto.ca, stark.draper@utoronto.ca).}% <-this % stops a space
%\thanks{Copyright (c) 2021 IEEE. Personal use of this material is permitted.  However, permission to use this material for any other purposes must be obtained from the IEEE by sending a request to pubs-permissions@ieee.org.}
}% <-this % stops a space}

% The paper headers
%%\markboth{Journal of \LaTeX\ Class Files,~Vol.~14, No.~8, August~2021}%
%%{Shell \MakeLowercase{\textit{et al.}}: A Sample Article Using IEEEtran.cls for IEEE Journals}

%\IEEEpubid{0000--0000/00\$00.00~\copyright~2021 IEEE}
% Remember, if you use this you must call \IEEEpubidadjcol in the second
% column for its text to clear the IEEEpubid mark.

\maketitle

\begin{abstract}
Coded distributed computing was recently introduced to mitigate the effect of stragglers on distributed computing. This paper combines ideas of approximate computing with coded computing to further accelerate computation. We propose {\em successive approximation coding} (SAC) techniques that realize a tradeoff between accuracy and speed, allowing the distributed computing system to produce approximations that increase in accuracy over time. If a sufficient number of compute nodes finish their tasks, SAC exactly recovers the desired computation. We theoretically provide design guidelines for our SAC techniques, and numerically show that SAC achieves a better accuracy-speed tradeoff in comparison with previous methods.
\end{abstract}

\section{Introduction}
Distributed computing is necessary for handling modern real-time data analytics and computationally intensive applications such as genome and DNA sequencing, and deep learning. To meet this need, distributed systems are increasing in complexity and size. This gives rise to novel challenges that plague today’s distributed computing systems. These include failures, stragglers, communication bottlenecks, and security and privacy issues. The emerging field of coded distributed computing (CDC) comprises a set of promising techniques that deal with these challenges~\cite{lee2017speeding, dutta2016short, lee2017high, yu2017polynomial, dutta2019optimal, kiani2018exploitation, kianidehkordi2020hierarchical, mallick2019rateless, yu2020straggler, dutta2018unified, yu2019lagrange}. CDC employs ideas from coding theory to form redundant computations such that only some sufficient number of computing nodes (called workers) need to complete their tasks before desired computation can be recovered. The number of required workers is called  the {\em recovery threshold}. In distributed computing systems that consist of tens or hundreds of workers (or even more for emerging Federated learning applications), recovery thresholds need to be large because of the tradeoff between recovery threshold and per-worker communication load. In large-scale systems, typical CDC strategies minimize expensive communication from each worker by increasing the recovery threshold. However, the larger recovery threshold is, the lower tolerance the system has for failures, stragglers, attackers, or colluding nodes. Lower tolerance can result in unnecessary delay, power consumption, and the waste of underlying hardware resources (such as CPUs and GPUs). That said, the fundamental limits on recovery threshold and per-worker communication load can be improved by relaxing requirements for exact recovery.% most CDC works to date  limit themselves to exactly recover the desired computation.% and thus underestimate their performance.

In this paper, we combine ideas from approximate computing with CDC, producing a sequence of approximations of our target computation that are increasingly accurate. In analogy with the recovery threshold, we introduce a new metric {\em approximate threshold} that quantifies the number of workers required for approximate recovery within some degree of error. We identify a tradeoff between the approximate threshold and the magnitude of the error and provide explicit CDC methods that illustrate this tradeoff. Our approximate CDC methods hold promise for many modern error-tolerant applications such as learning deep neural network models. In these applications, approximated results are acceptable and often inevitable. In the remainder of this introductory section, we first overview prior CDC schemes to provide the proper context to present our contributions. We then detail our contributions and outline the remainder of the paper.

%In this paper, we combine ideas from approximate computing with CDC, relaxing requirements for exact recovery and producing a sequence of approximate computation of increasing accuracy. In analogy with the recovery threshold, we introduce a new metric {\em approximate threshold} that quantifies number of workers required for approximate recovery within some degree of error. We identify a fundamental tradeoff between the approximate threshold and the magnitude of the error and provide explicit CDC constructions that illustrate this tradeoff. Our approximate CDC constructions can be used in many modern error-tolerant applications such as learning deep neural network models. In these applications, approximated results are acceptable and often inevitable. In the remainder of this introductory section, we overview prior CDC schemes that we build on and then present our contributions.

\subsection{Background}
\subsubsection{Polynomial-based CDC}
 The foundation of CDC schemes is built on novel methods for the distributed multiplication of a pair of matrices $A$ and $B$ to form their product $AB$. Probably the largest group of CDC schemes designed for distributed matrix multiplication are based on polynomials. In all, a pair of {\em encoding} polynomials are used separately to encode the $A$ and $B $ matrices. Computing the product of these encoding polynomials results in another polynomial, called the {\em decoding} polynomial. The degree of the decoding polynomial equals the sum of the degrees of the encoding polynomials. Each worker is tasked with multiplying two encoding polynomials, both of which are evaluated at the same point. Each worker's evaluation point is distinct from other workers' evaluation points. The completion of a worker's task is equivalent to calculating the decoding polynomial at that worker's evaluation point. Once a sufficient number of workers complete their tasks, evaluations of the decoding polynomial are known at a sufficient number of points that polynomial interpolation can be applied and the decoding polynomial can be fully recovered. The desired matrix product $AB$ is recovered from the interpolated decoding polynomial. 

We group polynomial-based CDC schemes based on how they recover the $AB$ product from the interpolated polynomial: {\em coefficient-based} or {\em point-based}. In the former group, the $AB$ product is equal to one (or more) of the coefficients in the decoding polynomial. In the latter, a post-decoding calculation is required wherein the decoding polynomial is interpolated at (new) points and the $AB$ product is then recovered by linearly combining the new interpolated results. For example, Polynomial codes~\cite{yu2017polynomial} are one of the earliest CDC constructions. They belong to the coefficient-based group because they recover the $AB$ product by concatenating all the coefficients of the decoding polynomial. Another coefficient-based CDC scheme is MatDot codes~\cite{dutta2019optimal}. The recovery threshold of Polynomial codes is reduced by MatDot codes, albeit at the cost of increased communication per worker. One of the coefficients in the decoding polynomial of MatDot codes is equal to the $AB$ product. In~\cite{dutta2019optimal, yu2020straggler, dutta2018unified}, other coefficient-based CDC schemes are introduced that generalize and unify Polynomial and MatDot codes, providing a tradeoff between recovery threshold and per-worker communication load. 

There are two features common to all these coefficient-based CDC constructions. The first is that both the encoding and decoding polynomials are expanded in the {\em monomial} basis, $1,x,x^2,\ldots$. The second is that the interpolation at the decoding phase requires the solution of a system of linear equations that involves a {\em Vandermonde} matrix with real-valued entries. Through Vandermonde-based interpolation, coefficient-based CDC schemes recover the $AB$ product directly from the interpolated coefficients of the monomial basis. However, Vandermonde matrices defined over reals are ill-conditioning. The condition number of real Vandermonde matrices grows exponentially in the dimension of the matrix~\cite{quarteroni2010numerical}. Ill-conditioning can lead to numerical problems when the inverse is taken to perform interpolation. In contrast, point-based CDC constructions are not limited to the monomial basis nor to Vandermonde-based interpolation. They can use better-conditioned bases but must extract the desired computation from their coefficients through post-decoding calculations. Despite the requirement of post-decoding calculations, the use of other (non-monomial) polynomial bases has benefits when compared to coefficient-based CDC constructions. For example, OrthoMatDot codes~\cite{fahim2021numerically} use an {\em orthonormal} basis and {\em Chebyshev-Vandermonde} interpolation. This solves the ill-conditioning issue of MatDot codes. Another point-based CDC method is Lagrange coding~\cite{yu2019lagrange} which uses the {\em Lagrange} polynomial basis. Using the Lagrange basis allows Lagrange codes to extend to more general multi-variate polynomial computing beyond matrix multiplication with guarantees of straggler resilience, security, and privacy. Lagrange codes use Vandermonde interpolation which can again lead to an ill-conditioned problem. In addition to MatDot codes,~\cite{fahim2021numerically} applies a Chebyshev-Vandermonde decoder to Lagrange codes in order to mitigate ill-conditioning. %We will detail some of the aforementioned code constructions that are the most relative to our work later in Sec.~\ref{sec:benchmark}. 

\subsubsection{Approximation in CDC}
While most CDC works to date limit themselves to exact recovery, some recent literature has combined approximation with earlier CDC schemes~\cite{charalambides2021approximate,chang2019random, ferdinand2016anytime, zhu2017sequential, jahani2021codedsketch, gupta2018oversketch, 9509407}. Motivated by~\cite{drineas2006fast}, references~\cite{charalambides2021approximate} and~\cite{ chang2019random} use different random-based sampling techniques first to compress $A$ and $B$ and then use MatDot codes to encode the data matrices. Due to the randomized compression of their input matrices, the recovery thresholds in both~\cite{charalambides2021approximate} and~\cite{ chang2019random} are reduced. This accelerates recovery while providing only an approximation of the desired computation. In both~\cite{charalambides2021approximate} and~\cite{ chang2019random}, the compressing and coding steps are separate. Therefore, to provide another estimate with lower error, the input matrices should be newly compressed and a new coding step need to be used for the new random samples of inputs. Another line of work~\cite{ferdinand2016anytime} and~\cite{ zhu2017sequential} brings the idea of approximate computing to the encoding and decoding steps. Using maximum distance separable (MDS) codes,~\cite{ferdinand2016anytime} and~\cite{ zhu2017sequential} guarantee accuracy improvement over time by intelligently prioritizing and allocating tasks among workers. These methods while effective, are applied only to matrix-vector multiplication. Their extension to matrix-matrix multiplication is non-trivial. 

Approximation methods for matrix-matrix multiplication are introduced in~\cite{jahani2021codedsketch} and~\cite{ gupta2018oversketch} based on sketching. In~\cite{jahani2021codedsketch} and~\cite{ gupta2018oversketch} randomized pre-compression via sketching is used to reduce the dimension of input matrices, thereby, reducing the recovery threshold. In contrast to the aforementioned random sampling-based techniques (\hspace{1sp}\cite{charalambides2021approximate, chang2019random}), references~\cite{jahani2021codedsketch} and~\cite{gupta2018oversketch} jointly design the sketching step together with the coding step to minimize the recovery threshold while ensuring the required computation accuracy. Such optimal recovery threshold is achieved only in a probabilistic manner. In other words, there exists a non-zero probability of failing to recover the computation to the required accuracy. For us, the most relevant approximate CDC construction, and the one that motivated our work, is {\em $\epsilon$-approximate} MatDot codes~\cite{9509407}. In~\cite{9509407}, polynomial approximation techniques add a layer of approximation to MatDot codes. Through this approximation layer, the recovery threshold is roughly halved and the approximated computation is recovered with probability one. More details on~\cite{9509407} are provided in Sec.~\ref{sec:benchmark}.% We next present our contributions.

\subsection{Summary of Contributions}
In this paper, we develop two novel approximate CDC methods. Each produces a sequence of increasingly accurate approximations of the desired calculation as more and more workers report in. We term this {\em successive approximation coding} (SAC) and call each approximation in the sequence a {\em resolution layer}. Our first method can be viewed as an extension of $\epsilon$-approximate MatDot codes~\cite{9509407}. While~\cite{9509407} allows for a single resolution layer prior to exact recovery, our designs allow multiple resolutions. This enables the distributed system to stop sooner, whenever the resolution of the matrix product is satisfactorily high. In our first method, we divide resolution layers into groups. While resolution increases across layers, these increases are most significant across groups and comparatively negligible within the layers of a group. Our second method extends approximation recovery to point-based polynomial codes, such as OrthoMatDot codes\cite{fahim2021numerically} and Lagrange codes~\cite{yu2019lagrange}. This method can be viewed as a limiting version of our first method where each group contains a single layer and accuracy improves every time an additional worker completes its task. We call our two methods ``group-wise'' and ``layer-wise'' SAC. Due to the grouping of the resolution layers in group-wise SAC, the approximate recovery  improves in discrete steps (one step per group) as workers report in. For layer-wise SAC, the improvements are more continuous. We theoretically design our SAC methods in such a way that their error of approximation is minimized. Our simulation results show that in comparison with $\epsilon$-approximate MatDot codes~\cite{9509407}, our methods require a lower number of workers for approximate recovery (i.e., a lower approximation threshold) to attain the same degree of error.

An outline for the rest of the paper is as follows. In Sec.~\ref{SEC:problem}, we present the problem formulation that we consider. In Sec.~\ref{sec:SAC} and~\ref{SEC:rsac}, we introduce our two SAC methods in detail. In Sec.~\ref{sec:simulation}, we experimentally show the benefits of each of our SAC methods. We conclude in Sec.~\ref{sec:conclusion}.

\section{Problem Formulation}\label{SEC:problem}
In this section, we first present the system model. We then discuss relevant performance measures. Finally, we detail previous CDC schemes that set the stage for our work.

\subsection{System Model}
We consider a distributed computing system that consists of a master and $N$ workers. The job is to compute the matrix-matrix product $AB$, where $A \in \mathbb{R}^{N_x\times N_z}$ and $B \in \mathbb{R}^{N_z\times N_y}$. We partition $A$ vertically into $K$ equally-sized submatrices $A=[A_1,\ldots,A_K]$, where for any $k\in[K]$ that $N_z/K \in \mathbb{Z}$, $A_k$ is a matrix of size $N_x\times N_z/K$\footnote{For any positive integer $n$, we use $[n]$ to refer to the set $\{1,\ldots,n\}$.}. Similarly, the $B$ matrix is partitioned horizontally into $K$ equally-sized submatrices $B^T=[B_1^T,\ldots,B_K^T]$, where $B_k$ is a $N_z/K\times N_y$ matrix, $k\in[K]$. Due to this partitioning, the $AB$ product can be calculated as a sum of $K$ outer products, $AB=\sum_{k=1}^K A_kB_k$. We use the notation $K$ to denote the {\em information dimension} of a code. The information dimension refers to the number of useful (non-redundant) computations
into which the main computational job is partitioned. For a fair comparison, we fix both the information dimension, $K$, and the number of workers, $N$, across different CDC schemes in this paper. %We aim to parallelize this computation across $N$ workers while enabling tradeoffs between the accuracy of results and the computation speed.

\subsection{Performance Metric} 
We use the following metrics throughout the paper to benchmark our proposed method against prior work. 

\begin{definition}[Recovery threshold] 
The recovery threshold is the number of workers required to complete their tasks for the master to be able to obtain the {\em exact} result. To denote the recovery threshold of a code, we use $R_{{\rm x}}$ with the subscript ${{\rm x}}$ indicating the type of the code. For example, for the codes that are used in this paper, their recovery thresholds are summarized in the second column of Table~\ref{table:terminology}. In CDC, $N-R$ can be thought of as the maximum number of stragglers that the distributed system can tolerate\footnote{We use the notations without their predetermined subscript code name in order to refer to a general code. For example, $R$ refers to the recovery threshold of a general CDC scheme, while $R_{\text{MD}}$ refers particularly to the recovery threshold of MatDot codes.}.
\end{definition}
\begin{table*}[h]
\caption{Summary of Parameters Used in This Paper for Different Coded Distributed Computing (CDC) Schemes. For Different CDC Schemes, the Information Dimension and the Number of Workers Are Set to Be $K$ and $N$, Respectively.}
\center
\begin{tabular}{|l|l|l|l|l|} \hline
CDC scheme & Recovery Thr. & $\#$ Resolution layers & Approx. Thr. & Rel. Err. \\ \hline
 MatDot (MD)~\cite{dutta2019optimal} & $R_{\text{MD}}=2K-1$ & 0 & $-$ & $-$ \\ \hline
 $\epsilon$-approximate MD~\cite{9509407} & $R_{\epsilon\text{AMD}}=R_{\text{MD}}$& 1 & $R_{\epsilon\text{AMD},1}=K$ & $\tau_{\epsilon\text{AMD},1}$ \\ \hline
 OrthoMatDot~\cite{fahim2021numerically} &$R_{\text{OMD}}=R_{\text{MD}}$ & 0 & $-$  & $-$  \\ \hline
 Lagrange~\cite{yu2019lagrange} & $R_{\text{Lag}}=R_{\text{MD}}$ & 0 & $-$  & $-$\\ \hline
 	Group-wise SAC & $R_{\text{G-SAC}}\geq R_{\text{MD}}$ & $L_{\text{G-SAC}}, \; {l\in [L_{\text{G-SAC}}]}$& $\{R_{\text{G-SAC},l}\}_l$ & $\{\tau_{\text{G-SAC},l}\}_l$  \\ \hline
 Layer-wise SAC & $R_{\text{L-SAC}}=R_{\text{MD}}$ & $L_{\text{L-SAC}}, \; {l\in [L_{\text{L-SAC}}]}$ & $\{R_{\text{L-SAC},l}\}_l$ & $\{\tau_{\text{L-SAC},l}\}_l$   \\ \hline
 \end{tabular}
 \label{table:terminology}
\end{table*}

While $R$ workers are required to complete their tasks in order to recover the $AB$ product exactly, the completion of a {\em smaller} number of workers can be sufficient to approximate the $AB$ product. For this approximation procedure, we next define a positive triple $(L_{{\rm x}},R_{{\rm x},l}, \tau_{{\rm x},l})$, for a CDC scheme of type-${{\rm x}}$. This triple specifies the number of resolution layers ($L_{{\rm x}}$), the approximate threshold ($R_{{\rm x},l}$), and the relative error ($\tau_{{\rm x},l}$) for the $l$th layer, where $l \in [L_{{\rm x}}]$.

\begin{definition}[Resolution layer]
 The resolution layer corresponds to a time index when a sufficient number of workers have completed their tasks so as to be able to generate an improved approximate of the $AB$ product. For a CDC scheme of type-${\rm x}$, the parameter $L_{{\rm x}}$ with a subscript of its code type is used to refer to the number of resolution layers. The parameter $L_{{\rm x}}$ ranges between some minimum and maximum admissible values which will be detailed in Sec.~\ref{sec:SAC} and~\ref{SEC:rsac}. We list a summary of this notation in the third column of Table~\ref{table:terminology}.
\end{definition}

\begin{definition}[Approximate threshold]
 The approximate threshold is the number of workers required to approximate the $AB$ product to some degree of accuracy. We denote the approximate threshold with $R_{{\rm x},l}$, where the first subscript ${\rm x}$ indicates the type of the code and $l$ indicates the index of the resolution layer\footnote{Note that the recovery threshold is an extreme case of approximation threshold when the error is 0, so we use the same letter, although the subscripting is different.}, which ranges from 1 to $L_{{\rm x}}$. A summary of notation used for the approximate threshold of different CDC schemes is provided in the fourth column of Table~\ref{table:terminology}.
\end{definition}

\begin{definition}[Relative error]
At layer $l \in [L_{{\rm x}}]$, if the matrix $\tilde{C}_l \in \mathbb{R}^{N_x\times N_y}$ computes the approximate of the $AB$ product, the relative error is defined as the Frobenius norm of the difference between $\tilde{C}_l$ and the exact $AB$ product, normalized by the Frobenius norm of the matrix $AB$, i.e., $ \|\tilde{C}_l -AB \|_F^2 / \|AB\|_F^2$. We use the notation $\tau_{{\rm x},l}$ with two subscripts (using similar logic  to that of the approximation threshold) to refer to the relative error. We list the notation corresponding to the relative error in the last column of Table~\ref{table:terminology}.
 \end{definition}

\begin{remark}
Throughout this paper, we use the terms ``exact''  and ``approximate'' to differentiate between cases where the exact recovery of the $AB$ product is possible or is not. We remark that in the situation of finite precision, even in the case of exact recovery, computation error is generally not identically zero due to numerical errors. Numerical error is of course not limited to exact recovery, it is also present in approximate recovery. In the next section, we detail a technique introduced in~\cite{fahim2021numerically} to reduce issues of numerical precision. We use similar techniques to implement our  methods in Sec.~\ref{sec:simulation}.
\end{remark}

\subsection{Related Work for Benchmarking}\label{sec:benchmark}
% We now detail some prior CDC schemes that are related to our work~\cite{dutta2019optimal, 9509407, fahim2021numerically, yu2019lagrange}. In all, a pair of {\em encoding} polynomials are used separately to encode $A$ and $B $. The product of these encoded data matrices results in another polynomial, called the {\em decoding} polynomial, of degree equals the sum of the degrees of the encoding polynomials. Each worker is tasked with multiplying two encoding polynomials, which have been evaluated at a certain point (distinct from other workers' evaluation points). The task's result is equivalent to the calculation of the decoding polynomial at the worker's evaluation point. Once a sufficient workers complete their tasks, the decoding polynomial evaluation is done at a sufficient number of points that polynomial interpolation can be applied and the decoding polynomial recovered. Finally, the master recovers the desired $AB$ product from the interpolated decoding polynomial. We group these CDC schemes based on how they recover the $AB$ product from the interpolated polynomial: coefficient-based or point-based. For the former, the $AB$ product is equal to the one of the coefficients in the interpolated polynomial. In the later, a post-decoding calculation is required. The interpolated polynomial  is evaluated at (new) points and the $AB$ product is then recovered by linearly combining the new evaluation results.   

\textbf{Coefficient-based polynomial CDC schemes:} MatDot codes~\cite{dutta2019optimal} encode $\{A_k\}_{k=1}^K$ and $\{B_k\}_{k=1}^K$ to generate encoding polynomials as $\hat{A}(x) = \sum_{k=1}^K A_kx^{k-1}$ and $\hat{B}(x) = \sum_{k=0}^{K-1} B_{K-k}x^{k}$. The multiplication of $\hat{A}(x_n)$ and $\hat{B}(x_n)$ is assigned to worker $n\in [N]$, for an arbitrary real value of $x_n$ (distinct from $x_m$ for $m\neq n$). The decoding polynomial $\hat{A}(x)\hat{B}(x)$ for $x \in \{x_1,\ldots,x_N\}$ is a $(2K-2)$-degree polynomial, where the coefficient of $x^{K-1}$ is equal to the target $AB$ product. The completion of any $2K-1$ calculations of encoded products suffices to recover the entire polynomial, including the $(K-1)$th coefficient. We use $R_{\text{MD}}=2K-1$ to refer to the recovery threshold of MatDot codes. Note that $R_{\text{MD}}\leq N$. The recovery process of MatDot codes is an exact one. Either the $AB$ product can be calculated exactly (once any $R_{\text{MD}}$ workers complete their tasks), or nothing can be said about $AB$ (when fewer than $R_{\text{MD}}$ workers have reported in). %To make this recovery process an approximated one or in other words to improve the recovery threshold at the cost of losing some accuracy, %we next review ``$\epsilon-$approximate'' MatDot codes~\cite{9509407} and introduce two novel methods on top of that.   
%\subsection{$\epsilon-$Approximated MatDot codes~\cite{9509407}}

Based on MatDot codes, $\epsilon$-approximate MatDot codes~\cite{9509407} introduce an approximate recovery process. The strategy yields an approximation of the $AB$ product when fewer than $2K-1$ workers have completed their tasks. In particular, when only $R_{\epsilon\text{AMD},1}=K$ workers have completed their tasks, the $AB$ product can be estimated with some error. The approximation is achieved by interpolating a $(K-1)$-degree polynomial $\hat{P}(x)$, where $\hat{P}(x)$ is the residual polynomial of the division of $\hat{A}(x)\hat{B}(x)$ by $x^K$. If we use $Q(x)$ to denote the quotient polynomial, then we can rewrite the decoding polynomial as $\hat{A}(x)\hat{B}(x) = x^K\hat{Q}(x)+\hat{P}(x)$, where $x^K\hat{Q}(x)$ consists of all higher-order terms and $\hat{P}(x)$ consists of all lower-order terms of $\hat{A}(x)\hat{B}(x)$ polynomial. Note that the leading coefficient (the coefficient of $x^{K-1}$) of $\hat{P}(x)$ is equal to the target $AB=\sum_{k=1}^K A_kB_k$ product and $\hat{A}(x)\hat{B}(x) \approx \hat{P}(x)$ if $x$ is sufficiently small. The recovery procedure can be improved by an exact calculation of $AB$; possible once any $R_{\epsilon\text{AMD}}=2K-1$ workers finish their tasks and the $(2K-2)$-degree polynomial $\hat{A}(x)\hat{B}(x)$ can be interpolated. 
%Subsequently, the exact $AB$ product can be recovered once any $R_{\epsilon\text{AMD}}=2K-1$ workers have completed their tasks
%\begin{remark}
%In coefficient-based polynomial CDC schemes, both the encoding and decoding polynomials are expanded in the {\em monomial} basis, i.e., $1,x,x^2,\ldots$ and their interpolation requires solving a system of linear equations that involves a {\em Vandermonde} matrix with real valued entries. Troughout this real Vandermonde-based interpolation, this group of CDC schemes do not need to any further post decoding calculation or in other words can recover the $AB$ product directly from the interpolated coefficients of the monomial basis. However, taking the inverse of the real Vandermonde matrix suffers from ill-conditioning: condition number of real Vandermonde matrices grows exponentially in the dimension of the matrix~\cite{fahim2021numerically, ramamoorthy2021numerically}.
%\end{remark}

\textbf{Point-based polynomial CDC schemes:} To address the problem of ill-conditioned Vandermonde matrices, OrthoMatDot codes are introduced in~\cite{fahim2021numerically}. The key idea of OrthoMatDot codes is to use an {\em orthonormal} basis instead of the ill-conditioned monomial basis used in MatDot codes. An orthonormal basis is a system of algebraic polynomials, $\{O_k(x)\}$, with the degree of $O_k(x)$ equal to $k$ for all $k=0,1,\ldots$. The polynomials are orthonormal on the interval $(-1,1)$ with respect to a weight measure $w(x)$~\cite{quarteroni2010numerical}. In other words, for same nonnegative integrable function $w(x)$ in $(-1,1)$, the inner products satisfy 
\begin{align*}
\int_{1}^{-1} O_k(x)O_m(x)w(x)dx=\begin{cases}1 & \text{if } k\neq m \\
0 & \text{otherwise} \end{cases}.
\end{align*}
For example, the Chebyshev polynomials are defined via a recursive formula~\cite{quarteroni2010numerical}
\begin{align*}
\begin{cases}p_{k+1}(x) = 2xp_k(x)-p_{k-1}(x) & k=1,2,\ldots \\
p_0(x)=1, \; p_1(x)=x & \end{cases}.
\end{align*}
The Chebyshev polynomials produce a well-known class of orthonormal polynomials; $O_0(x)=\frac{1}{\sqrt{2}}p_0(x), O_k(x)=p_k(x)$, for $k=1,2,\ldots$ and $w(x)=\frac{2}{\pi \sqrt{1-x^2}}$. Using such an orthonormal polynomials (e.g., the Chebyshev polynomials), OrthoMatDot codes generate $N$ pairs of encoding polynomials $(\tilde{O}_A(x_n),\tilde{O}_B(x_n))$, $n\in [N]$, from the $2K$ matrices $A_1,\ldots,A_K$ and $B_1,\ldots,B_K$. The encoding polynomials are $\tilde{O}_A(x)=\sum_{k=1}^K A_kO_k(x)$ and $\tilde{O}_B(x)=\sum_{k=1}^K B_kO_k(x)$. As before, worker $n\in [N]$ is tasked with the multiplication of $\tilde{O}_A(x_n)$ and $\tilde{O}_B(x_n)$, where $\{x_1,\ldots,x_N\}$ is a set of distinct real evaluation points. Since the decoding polynomial $\tilde{O}_A(x)\tilde{O}_B(x)$ is, in general, a $2K-2$ degree polynomial, the recovery threshold of OrthoMatDot codes is equal to $R_{\text{OMD}}=2K-1$. Assuming that the workers indexed by $j_1,\ldots,j_{2K-1}$ are the first to complete among all the $N$ workers, the decoding procedure of OrthoMatDot codes involves a {\em Chebyshev-Vandermonde} matrix, defined as
\begin{equation*}
\begin{pmatrix}
O_0(x_{j_1}) & \cdots & O_0(x_{j_{2K-1}}) \\
\vdots & \ddots & \vdots \\
O_{2K-2}(x_{j_1}) & \cdots & O_{2K-2}(x_{j_{2K-1}})
\end{pmatrix}.
\end{equation*}
A careful choice of evaluation points, e.g., $x_n=\eta_n^{(N)}$, where $\{\eta_n^{(N)}\}$ are the $N$ (distinct real) roots of $O_N(x)$, ensures that the Chebyshev Vandermonde system is well conditioned~\cite{fahim2021numerically}. While well-conditioned, OrthoMatDot codes do require a post-decoding calculation in order to recover $AB$. Because the orthonormal basis is used,~\cite{fahim2021numerically} proves that the $AB$ product can be recovered via the sum $\sum_{k=1}^K \frac{2}{K}\tilde{O}_A(\eta_k^{(K)})\tilde{O}_B(\eta_k^{(K)})$, where the $\eta_k^{(K)}$ are the roots of $O_K(x)$.

Lagrange codes are another point-based polynomial CDC scheme originally introduced in~\cite{yu2019lagrange}. Lagrange codes compute a general multivariate polynomial. In the following, we show how to apply Lagrange codes to our particular matrix multiplication problem. Lagrange codes encode data matrices $A_1,\ldots,A_K$ to generate the encoding polynomial as $\tilde{L}_A(x)=\sum_{k=1}^K A_kL_k(x)$, where $\{L_k(x)\}$ is Lagrange basis. Here, $L_k(x)$ is a $(K-1)$-degree polynomial, defined as $L_k(x)=\prod_{j\neq k}\frac{(x-y_j)}{(y_k-y_j)}$. Similarly, the encoding polynomial corresponding to data matrices $B_1,\ldots,B_K$ is $\tilde{L}_B(x)=\sum_{k=1}^K B_kL_k(x)$. Note that $\tilde{L}_A(y_k) = A_k$ and $\tilde{L}_B(y_k)=B_k$. The master assigns the multiplication of $\tilde{L}_A(x_n)$ and $\tilde{L}_B(x_n)$ to worker $n\in [N]$, where $x_1,\ldots,x_N$ are $N$ distinct real numbers (possibly distinct from $\{y_k\}_{k\in [K]}$). Since each Lagrange polynomial $L_k(x)$ is of degree $K-1$, the recovery threshold of Lagrange codes is $R_{\text{Lag}}=2K-1$. When the polynomial $\tilde{L}_A(x)\tilde{L}_B(x)$ is ready for interpolation, Lagrange codes~\cite{yu2019lagrange} solve a real-valued Vandermonde system of equations. After decoding, the $\tilde{L}_A(x)\tilde{L}_B(x)$ polynomial needs to be interpolated at $K$ distinct interpolation points $y_1,\ldots,y_K$ and then summed together to recover the $AB$ product, $AB = \sum_{k=1}^K \tilde{L}_A(y_k)\tilde{L}_B(y_k)$. 

%\begin{remark}
%Expanding the encoding polynomials either in an Orthonormal or in a Lagrange polynomial basis allows the point-based polynomial CDC schemes to expand their decoding polynomials in Orthonormal polynomial basis and thus enables the well-conditioned Chebyshev-Vandermonde interpolation. %Despite this, the point-based polynomial CDC schemes just similar to coefficient-based polynomial CDC schemes can yet expand their decoding polynomials in monomial basis and suffer from an ill-conditioned problem because of the real Vandermonde-based decoding procedure. 
%\end{remark}

\begin{remark} Among all above CDC schemes~\cite{dutta2019optimal, 9509407, fahim2021numerically, yu2019lagrange}, only $\epsilon$-MatDot codes have a single resolution layer, i.e., $L_{\epsilon\text{AMD}}=1$. All others support only an exact recovery process. They thus have no resolution layer.
\end{remark}
\section{Group-wise Successive Approximated Coding}\label{sec:SAC}
In this section, we introduce group-wise SAC. Our method extends the single-layer approximation procedure of $\epsilon$-approximate MatDot codes~\cite{9509407} to multiple resolution layers. The recovery process of group-wise SAC consists of multiple resolution layers before exact recovery. This enables successive improvements of the approximations of the $AB$ product as additional workers report in. While such successive improvements are (marginally) obtained whenever an additional resolution layer completes, the major improvements of group-wise SAC occur once groups of layers complete. We next detail our constructions.

In group-wise SAC, we divide the resolution layers into $D$ (disjoint and consecutive) groups. Group $d \in [D]$ contains $L_d$ layers. The total number of layers is $ L_{\text{G-SAC}}=\sum_{d=1}^D L_d$. The approximation of the $AB$ product that is obtained in layer $l_{d,i}$, $d \in [D]$ and $i \in [L_d]$, improves as either $d$ or $i$ increases. Resolution increases in discrete steps when $d$ increments but only slightly as $l$ increments. To explain better, consider the first resolution layer of the $d$th group (layer $l_{d,1}=1+\sum_{j=1}^{d-1} L_j$). This layer provides an approximation of significantly higher resolution than earlier layers (layers in the $(d-1)$th group). On the other hand, in the same group $d$, layer $l_{d,i}$ approximates $AB$ only slightly better than the previous layer, $l_{d,i-1}$. We next provide the approximate threshold required for each of these resolution layers.

In layer $l \in [L_{\text{G-SAC}}]$, the approximate threshold is $R_{\text{G-SAC},l}$. I.e., our method approximates the $AB$ product when $R_{\text{G-SAC},l}$ workers have completed their tasks. While in $\epsilon$-approximate MatDot codes~\cite{9509407} the approximate threshold is $K$, our group-wise SAC allows the approximate threshold of the first layer to be any arbitrary (integer) value less than or equal to $K$, i.e., $R_{\text{G-SAC},1} \leq K$. As more workers complete tasks, the initial resolution can be improved. Across resolution layers $l\in \{2,3,\ldots, L_{\text{G-SAC}}\}$, our method is designed in such a way that the approximate thresholds satisfy $R_{\text{G-SAC},l}= R_{\text{G-SAC},l-1} + 1$ (cf. Secs.~\ref{sec:SAMD} and~\ref{sec:GSAMD} for details of this design). Finally, our method recovers the exact $AB$ product when a sufficient number of workers (larger than or equal to $2K-1$) complete their tasks. In other words, the recovery threshold of our group-wise SAC satisfies $R_{\text{G-SAC}} \geq 2K-1$. Therefore, one can conclude that $L_{\text{G-SAC}}\in \{R_{\text{G-SAC}}-K,\ldots, R_{\text{G-SAC}}-1\}$ (see App.~\ref{app:Lrange} for proof). Next, to develop the basic ideas in Sec.~\ref{sec:SAMD}, we detail group-wise SAC for  $D=2$ groups. We call this {\em two-group} SAC. We generalize to {\em multi-group} SAC in Sec.~\ref{sec:GSAMD}.

\subsection{Two-Group SAC}\label{sec:SAMD}

We now introduce two-group SAC. We use MatDot codes~\cite{dutta2019optimal} to generate encoding polynomials $\hat{S}_A(x)$ and $\hat{S}_B(x)$ for two-group SAC. However, the two $(K-1)$-degree polynomials $\hat{S}_A(x)$ and $\hat{S}_B(x)$ are different from the encoding polynomials $\hat{A}(x)$ and $\hat{B}(x)$, used in the $\epsilon-$approximate MatDot code design detailed in Sec.~\ref{sec:benchmark}. The coefficients of $\hat{S}_A(x)$ and $\hat{S}_B(x)$ are the permuted version of the coefficients of $\hat{A}(x)$ and $\hat{B}(x)$, respectively. To permute the coefficients, we {\em uniformly} shuffle the pairs of submatrices $(A_1,B_1),\ldots,(A_K,B_K)$ to get $(A_{i_1},B_{i_1}),\ldots,(A_{i_K},B_{i_K})$ and then divide the latter into two groups. The first group consists of the first $K_1=R_{\text{G-SAC},1}$ pairs $(A_{i_1},B_{i_1}),\ldots,(A_{i_{K_1}},B_{i_{K_1}})$. The other $K_2=K-K_1$ pairs belong to the second group. We generate encoding polynomials as 
\begin{align*}
\hat{S}_A(x) &= \left( \sum_{k=1}^{K}A_{i_k}x^{k-1}\right) \; \text{and} \\ \hat{S}_B(x) &= \left(\sum_{k=0}^{K_1-1} B_{i_{K_1-k}}x^{k}\right) + \left(\sum_{k=0}^{K_2-1} B_{i_{K-k}}x^{K_1+k}\right) .
\end{align*}
%\begin{align*}
%\hat{S}_A(x) &= \left( \sum_{k=1}^{K}A_{i_k}x^{k-1}\right) \; \text{and} \\ \hat{S}_B(x) &= \left(\sum_{k=0}^{K_1-1} B_{i_{K_1-k}}x^{k}\right) + \left(\sum_{k=0}^{K_2-1} B_{i_{K-k}}x^{K_1+k}\right) .
%\end{align*}
% TWO COL

%Note that in our two-group SAC, the approximate threshold increments one-by-one, satisfying in $R_{\text{G-SAC},l}=R_{\text{G-SAC},1} + l - 1$. Also, our method recovers the exact $AB$ product when any $R_{\text{G-SAC}}=2K-1$ tasks are completed. From this, one can conclude that $L_{\text{G-SAC},1}\in \{K-1,\ldots, 2K-2\}$ (see App.~\ref{app:Lrange}), and thus the degree of polynomial $\hat{P}_l(x)$ would be at most for $l=$.
%In the resolution layer $l\in \{2,3,\ldots, L_{\text{S$\epsilon$AMD}}\}$, the approximate threshold equals $R_{\text{S$\epsilon$AMD},l}=R_{\text{S$\epsilon$AMD},1} + l - 1$. Similar to $\epsilon$-approximate MatDot codes, our method recover the exact $AB$ product when any $R_{\text{S$\epsilon$AMD}}=2K-1$ tasks are completed. From this, one can easily conclude that $L_{\text{S$\epsilon$AMD},1}\in \{K-1,\ldots, 2K-2\}$ (see App.~\ref{app:Lrange}).
% Note that the degree of polynomial $\hat{P}_l(x)$ must be smaller than the degree of $\hat{S}_A(x)\hat{S}_B(x)$. Thus, 

The master assigns the multiplication of $\hat{S}_A(x_n)$ and $\hat{S}_B(x_n)$ to worker $n\in[N]$. Once sufficient numbers of workers complete their tasks, the master first approximates the $AB$ product through $L_{\text{G-SAC}}=L_1 + L_2$ resolution layers and finally recovers the exact $AB$. This can be explained
as follows. Note that in two-group SAC, $L_1=K$, $L_2=K_2-1$ if $K_2>0$ and equals 0 otherwise. We also note that the matrix product $\hat{S}_A(x)\hat{S}_B(x)$ is a $(2K-2)$-degree polynomial. In the $l$th resolution layer, $l \in [L_{\text{G-SAC}}]$, $\hat{S}_A(x)\hat{S}_B(x)$ can be written as a sum of a $(l+K_1-2)$-degree polynomial $\hat{P}_l(x)$ and the polynomial $x^{l+K_1-1}\hat{Q}_l(x)$ of higher-order terms. I.e., $\hat{S}_A(x)\hat{S}_B(x)=x^{l+K_1-1}\hat{Q}_l(x) + \hat{P}_l(x)$. Similar to $\epsilon$-approximate MatDot codes, by setting the evaluation points $x_1,\ldots,x_N$ to be distinct and (sufficiently) small, we are able to approximate $\hat{P}_l(x_n)$ by $\hat{S}_A(x_n)\hat{S}_B(x_n)$ for any $n\in [N]$. We next detail the basic ideas of how to approximate $AB$ starting from the first resolution layer. 

%The master assigns the multiplication of $\hat{S}_A(x_n)$ and $\hat{S}_B(x_n)$ to worker $n\in[N]$. The matrix product $\hat{S}_A(x)\hat{S}_B(x)$ is a $(2K-2)$-degree polynomial. For decoding at resolution layer $l \in [L_{\text{G-SAC}}]$, $\hat{S}_A(x)\hat{S}_B(x)$ can be written as a summation of a $(K_1-1)$-degree polynomial $\hat{P}_1(x)$ and the polynomial $x^{K_1}\hat{Q}_1(x)$, which consists of higher-order terms. I.e., $\hat{S}_A(x)\hat{S}_B(x)=x^{K_1}\hat{Q}_1(x) + \hat{P}_1(x)$. Note that the coefficients in $\hat{P}_1(x)$ belong to the first $K_1$ pairs. Similar to $\epsilon$-approximate MatDot codes, by setting the evaluation points $x_1,\ldots,x_N$ to be distinct and (sufficiently) small, we are able to approximate $\hat{P}_1(x_n)$ by $\hat{S}_A(x_n)\hat{S}_B(x_n)$ for any $n\in [N]$.  We next detail this approximation procedure.

In the first resolution layer, once the fastest $R_{\text{G-SAC},1}=K_1$ workers complete their tasks, the master {\em approximately} interpolates $\hat{P}_1(x)$ to recover the leading coefficient, $\sum_{k=1}^{K_1}A_{i_k}B_{i_k}$, to some accuracy. Define $C: =AB$ and ${C}_{l}: =\sum_{k=1}^{m_l} A_{i_k}B_{i_k}$. For the first layer, $l=1$ and $m_1=K_1$. In the following theorem, we show how optimally to select a scaling $\beta$ such that $\beta {C}_{l}$ is a good estimate of $C$. We start from the {\em expected approximation error}, defined as $\mathbb{E}\left(\|C-\beta {C}_{l}\|_F^2 \right)$, where the expectation is taken with respect to the indices of the random permutation of the $(A_i,B_i)$ pairs. The proof of the following theorem is provided in App.~\ref{app:thm0}. 
\begin{theorem}\label{thm0}
For a uniform random permutation of the $(A_i,B_i)$ pairs, the optimum solution to $\operatorname*{argmin}_\beta \mathbb{E}\left(\|C-\beta {C}_l\|_F^2 \right)$ is 
\begin{align}\label{eq:thm0beta}
\beta^* = \frac{M_1 + 2  M_2}{M_1 + 2 \frac{(m_l - 1)}{(K-1)} M_2},
\end{align}  
where
\begin{align*}
M_1 = \sum_{i=1}^{K} \|A_iB_i\|_F^2 \; \text{and }
M_2 = \sum_{i,j=1 , i< j}^K \Tr\left( \left(A_iB_i\right)^T \left(A_jB_j \right)\right).
\end{align*}
%\begin{align*}
%M_1 &= \sum_{k=1}^{K} \|A_kB_k\|_F^2 \;\; \text{and}\\
%M_2 &= \sum_{i,j=1 , i\neq j}^K \Tr\left( \left(A_iB_i\right)^T \left(A_jB_j \right)\right).
%\end{align*}
% TWO COL
%M_2 &= \sum_{\substack{i,j=1 \\ i\neq j}}^K \Tr\left( \left(A_iB_i\right)^T \left(A_jB_j \right)\right).
%\begin{align}\label{opt1}
%\operatorname*{argmin}_\beta \mathbb{E}\left(\|C-\beta \hat{C}\|_F^2 \right).
%\end{align}
 \end{theorem} 
 \begin{remark}
 We note that in general, neither $M_1$ nor $M_2$ will be known because they depend on the $A_kB_k$ products which are the constituent computations we want. We note that while setting $\beta=\frac{K}{m_l}$ makes $\beta C_l$ an unbiased estimate of $C$ (see Eq.~(\ref{app:eq:thm1}) in App.~\ref{app:thm0}), that choice for $\beta$ does not necessarily minimize the expected approximation error.
 \end{remark}
\begin{remark}\label{corol1}
While the optimal choice of $\beta$ will not in general be known, one can approximate it in the following two cases:
\begin{itemize}%[noitemsep, topsep=0pt]
\item \textit{Case 1:} The $\beta^*$ is approximately equal to $1$ if $M_2$ is small compared to $M_1$. While in general we cannot check whether or not this condition holds, prior knowledge of the distributions of $A_i$ and $B_i$ matrices can help us to conclude when this condition will be likely to hold. For example, if the entries of $A_i, A_j, B_i$ and $B_j$ are independent and identically distributed (i.i.d) random variables of zero mean and high variance then, in expectation, $\Tr\left((A_iB_i)^T(A_jB_j)\right)$ is zero if $i\neq j$ and is nonzero if $i=j$. This implies that $\mathbb{E}( M_2)=0$ and $\mathbb{E}(M_1)\neq 0$, and thus $\beta^*$ is highly likely to be close to unity.       
\item \textit{Case 2:} In another extreme, if $M_1\ll M_2$, then the optimal $\beta^* \approx \frac{K-1}{m_l-1}$. In this case, the optimal $\beta^*$ is close to $\frac{K}{m_l}$ which per the earlier remark is an unbiased estimate. Similar to Case~1, the condition $M_1\ll M_2$ is quite likely to hold in certain situations. For example, consider a situation where the entries $(A_i)_{e,k}$ and $(A_j)_{e,k}$ in any two matrices $A_i$ and $A_j$, $i\neq j$, are strongly and positively correlated, and the entries of $B_i$ and $B_j$ are also strongly and positively correlated. If the entries of the $A_i$ and $B_i$ are independent and have zero mean, then $\mathbb{E}( M_2)$ is a sum of ${K \choose 2}$ large positive terms, while $\mathbb{E} (M_1)$ is a sum of only $K$ terms. Since the former has $\mathcal{O}(K)$ more large terms than the later, with a high probability $M_1 \ll M_2$ and $\beta^* \approx \frac{K-1}{m_l-1}$.
\end{itemize} 
\end{remark}

%Now, we explain how to estimate $AB$ in other resolution layers, beyond the first layer. In the $l$th resolution layer, $l \in [L_{\text{G-SAC}}]$, the matrix product $\hat{S}_A(x)\hat{S}_B(x)$ is rewritten as the sum of a $(l+K_1-2)$-degree polynomial $\hat{P}_l(x)$ and the polynomial $x^{l+K_1-1}\hat{Q}_l(x)$. In other words, $\hat{S}_A(x)\hat{S}_B(x)=x^{l+K_1-1}\hat{Q}_l(x) + \hat{P}_l(x)$, where $l\in [L_{\text{G-SAC}}]$. When $R_{\text{G-SAC},l}=l+K_1-1$ workers complete their tasks, the polynomial $\hat{P}_l(x)$ can be interpolated to some accuracy. The coefficient of $x^{K_1-1}$ in $\hat{P}_l(x)$ is equal to $\sum_{k=1}^{K_1} A_{i_k}B_{i_k}$ which can be used to approximate $AB$. 

Now, we generalize the approximation procedure of $AB$ to any resolution layer $l \in [L_{\text{G-SAC}}]$. In layer $l \in [L_{\text{G-SAC}}]$, when $R_{\text{G-SAC},l}=l+K_1-1$ workers complete their tasks, the polynomial $\hat{P}_l(x)$ can be interpolated to some accuracy. The coefficient of $x^{K_1-1}$ in $\hat{P}_l(x)$ is equal to $\sum_{k=1}^{K_1} A_{i_k}B_{i_k}$ which can be used to approximate $AB$. When $K_1 < K$, $\sum_{k=1}^{K_1} A_{i_k}B_{i_k}$ is the sum of only a subset of outer products rather than all the outer products. Using the same notation as Thm.~\ref{thm0}, we use $C_l$ to denote the partial sum $\sum_{k=1}^{K_1} A_{i_k}B_{i_k}$ and make $m_l$ equal to $K_1$ for the first $L_1$ resolution layers (i.e., for $l \in [L_1]$). Note that $\hat{S}_A(x)\hat{S}_B(x)$ estimates $\hat{P}_l(x)$ more accurately than does $\hat{P}_{l-1}(x)$ because $\hat{P}_l(x)$ contains more terms of the $\hat{S}_A(x)\hat{S}_B(x)$ polynomial. This leads to the gradual improvement (in expectation) in the resolution of $C_l$ as $l$ increases from 1 to $L_1$. 

If $l$ exceeds $L_1$, either the exact recovery is enabled (if $L_2=0$) or the resolution layer $l=L_1+1$ approaches (if $L_2>0$). In either case, $C_{L_1+1}=\sum_{k=1}^{K} A_{i_k}B_{i_k}$ can be computed to some accuracy which yields a strictly better resolution of the full
summation. In the case of $L_2>0$, as $l$ increases from $L_1+1$ to $L_{\text{G-SAC}}$, the resolution of $C_l=\sum_{k=1}^{K} A_{i_k}B_{i_k}$ is gradually improved (in expectation). When $K_1 = K$, our method is similar to $\epsilon$-approximate MatDot codes in the sense that the master needs to wait for the same number of workers ($R_{\text{G-SAC},1}=K$) to provide the first approximation of $AB$. However, in this case, our method still outperforms $\epsilon$-approximate coding to some degree because the master can gradually improve on this resolution prior to exact recovery as $l$ increases from 1 to $L_{\text{G-SAC}}$.

Lastly, when any $R_{\text{G-SAC}}=2K-1$ workers complete their tasks, the master is able to interpolate the $\hat{S}_A(x)\hat{S}_B(x)$ polynomial exactly and recover its coefficients. When $K_1<K$, the coefficient of $x^{K_1-1}$ is equal to $\sum_{k=1}^{K_1} A_{i_k}B_{i_k}$ and the coefficient of $x^{K+K_1-1}$ is equal to $\sum_{k=K_1+1}^{K} A_{i_k}B_{i_k}$. The summation of these two coefficients yields the $AB$ product without approximation error. When $K_1=K$, only the coefficient of the $x^{K_1-1}$ term is required because it is equal to the full summation $\sum_{k=1}^{K} A_{i_k}B_{i_k}$.

\begin{figure*}[ht]
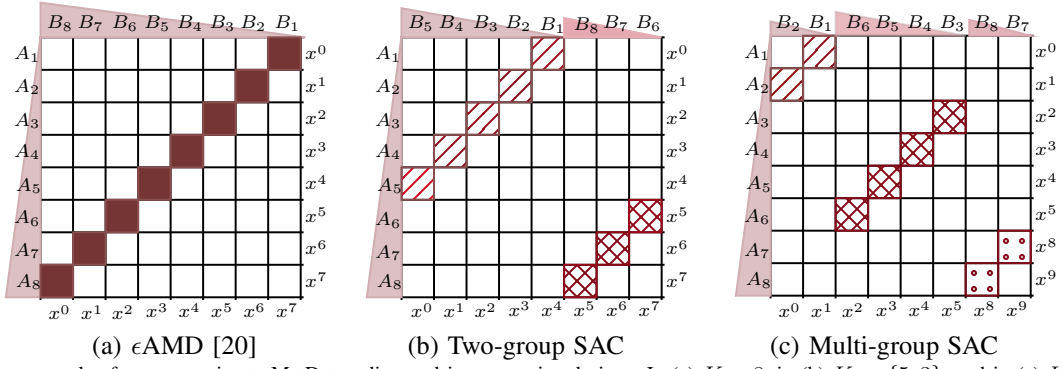

\centering
\include{./image/doubleEapprox}
  \vspace{-5ex}
\caption{Motivating examples for $\epsilon$-approximate MatDot coding and its successive designs. In (a) $K=8$, in (b) $K_i \in \{5,3\} $, and in (c) $K_i\in \{2,4,2\}$.} \label{fig:motiveexp}
\vspace{-3ex}
\end{figure*}

\begin{example}[Two-group SAC] \label{exp:samd}
 In Fig.~\ref{fig:motiveexp}(a,b), a motivating example benchmarks our two-group SAC against $\epsilon$-approximate MatDot codes~\cite{9509407} ($\epsilon$AMD), where $K=8$. Each subfigure is an array that consists of $8$ rows and $8$ columns. In Fig.~\ref{fig:motiveexp}(a) which presents the $\epsilon$AMD codes, each row $i\in [8]$ corresponds to the coefficient of $x^{i-1}$ in the encoding polynomials $\hat{A}(x)$. This coefficient is assumed to be equal to the submatrix $A_i$. Similarly, the $i$th column corresponds to the $B_{9-i}$ submatrix. As a result, the $(i,j)$th intersection corresponds to the product of the $i$th and $j$th coefficients in $\hat{A}(x)$ and $\hat{B}(x)$, respectively. Note that the arrays consist of $15$ antidiagonals (a diagonal from the top right to the bottom left). The summation of all coefficients along on each antidiagonal is equal to one of the coefficients in the decoding polynomial. For example, in Fig.~\ref{fig:motiveexp}(a), the sum of coefficients in the 8 solid boxes of the main antidiagonal is the coefficient of $x^{7}$ in $\hat{A}(x)\hat{B}(x)$, equal to $AB=\sum_{i=1}^8 A_iB_i$. 

 On the other hand, Figure~\ref{fig:motiveexp}(b) presents the two-layer SAC where the $i$th column corresponds to $B_{6-i}$ if $i<6$ and corresponds to $B_{14-i}$ otherwise. This can be justified via the difference in the encoding procedures of two-group SAC and $\epsilon$AMD which is due to the coefficient of $x^{i-1}$, $i \in [8]$, in $\hat{S}_B(x)$ and $\hat{B}(x)$. In Fig.~\ref{fig:motiveexp}(b), the summation of coefficients in the hatched boxes is the coefficient of $x^{4}$ in the $\hat{S}_A(x)\hat{S}_B(x)$ polynomial. This is equal to $\sum_{i=1}^5 A_iB_i\approx AB$. The summation of coefficients in the grid boxes  is the coefficient of $x^{12}$, equal to $\sum_{i=6}^8 A_iB_i$. This can be summed with $\sum_{i=1}^5 A_iB_i$ to recover the exact $AB$ product. 
 
 While in Fig.~\ref{fig:motiveexp}(a), the master needs to wait for $R_{\text{$\epsilon$AMD},1}=8$ completed workers to provide an estimate of $AB$, in Fig.~\ref{fig:motiveexp}(b), the master only needs to wait for $R_{\text{G-SAC},1}=5$ workers to provide an initial relatively low resolution. It then can gradually improve this resolution as $l$ increases from 1 to 8. Once $R_{\text{G-SAC},9}=13$ workers have completed their tasks, the master provides another relatively high resolution. It can improve this resolution slightly at $l=L_{\text{G-SAC}}=10$. In both Figs.~\ref{fig:motiveexp}(a,b), the master recovers the exact $AB$ product, once any $R_{\text{$\epsilon$AMD}}=R_{\text{G-SAC}}=15$ workers complete.
 \end{example} 

\subsection{Multi-Group SAC}\label{sec:GSAMD}
We now generalize two-group SAC to multiple groups. 
Similar to two-group SAC, we uniformly shuffle the pairs of submatrices so that $i_1,\ldots,i_K$ refer to the indices of the shuffled pairs, i.e., $(A_{i_1},B_{i_1}),\ldots,(A_{i_K},B_{i_K})$. As noted before, the reason that we use uniform shuffling is to provide improved estimate of $AB$ (in expectation) as more workers report in. We divide these shuffled pairs into $D\geq 2$ groups. The $d$th group consists of $K_d$ pairs $(A_{i_{k+\sum_{j=1}^{d-1}K_j}},B_{i_{k+\sum_{j=1}^{d-1}K_j}})$, $k\in [K_d]$. Thus, $\sum_{d=1}^D K_d = K$. For simplicity of notation, we use $(A_{k}^{(d)},B_{k}^{(d)})$ to refer to the pair $(A_{i_{k+\sum_{j=1}^{d-1}K_j}},B_{i_{k+\sum_{j=1}^{d-1}K_j}})$, for all $k\in[K_d]$ and $d\in [D]$. We encode matrices as 
\begin{align*}
\hat{S}_{A}(x) &= \left(\sum_{k=1}^{K_1} A_k^{(1)} x^{k-1}\right) + x^{K_1} \left( \sum_{k=1}^{K_2} A_k^{(2)} x^{k-1}\right) \\ &+  \sum_{d=3}^{D} \left( x^{g(d)} \left(\sum_{k=1}^{K_d} A_k^{(d)} x^{k-1}\right)\right) \; \text{and} \\
\hat{S}_{B}(x)&= \left(\sum_{k=0}^{K_1-1} B_{K_1-k}^{(1)}x^{k}\right)+ x^{K_1}\left( \sum_{k=0}^{K_2-1} B_{K_2-k}^{(2)}x^{k}\right) \\ &+ \sum_{d=3}^D \left(x^{g(d)}\left( \sum_{k=0}^{K_d-1}B_{K_d-k}^{(d)}x^{k}\right)\right),
\end{align*}
%\begin{align*}
%\hat{S}_{A}(x)&= \left(\sum_{k=1}^{K_1} A_k^{(1)} x^{k-1}\right) + x^{K_1} \left( \sum_{k=1}^{K_2} A_k^{(2)} x^{k-1}\right) \\ &+  \sum_{d=3}^{D} \left( x^{g(d)} \left(\sum_{k=1}^{K_d} A_k^{(d)} x^{k-1}\right)\right) \; \text{and} \\
%\hat{S}_{B}(x)&= \left(\sum_{k=0}^{K_1-1} B_{K_1-k}^{(1)}x^{k}\right)+ x^{K_1}\left( \sum_{k=0}^{K_2-1} B_{K_2-k}^{(2)}x^{k}\right) \\ &+ \sum_{d=3}^D \left(x^{g(d)}\left( \sum_{k=0}^{K_d-1}B_{K_d-k}^{(d)}x^{k}\right)\right),
%\end{align*}
% two col
where $g(d)=K_1+\left(\sum_{j=3}^d \left(\left(\sum_{i=1}^{j-2} 2^{j-i-2} K_i\right) + K_{j-1}\right)\right) $ for $d\geq 3$. Next, worker $n\in[N]$ multiplies $\hat{S}_A(x_n)$ and $\hat{S}_B(x_n)$ for sufficiently small value of $x_n$ ($x_n$ and $x_{n'}$ are distinct if $n\neq n'$). The decoding polynomial $\hat{S}_A(x)\hat{S}_B(x)$ has degree $(\sum_{d=1}^D 2^{D-d} K_d)+K_D-2$. Therefore, in the exact recovery layer, the recovery threshold is equal to $R_{\text{G-SAC}}=(\sum_{d=1}^D 2^{D-d} K_d)+K_D-1$. If $D=2$, this recovery threshold is equal to that of two-group SAC, while if $D>2$, $R_{\text{G-SAC}} > 2K-1$ (cf. App.~\ref{app:RT} for proofs of these claims). In group  $d \in [D]$, whenever any $R_{\text{G-SAC},l_{d,1}}=\sum_{i=1}^d 2^{d-i} K_i$ workers complete their tasks, the master can interpolate the $(R_{\text{G-SAC},l_{d,1}}-1)-$degree polynomial $\hat{P}_{l_{d,1}}(x)$. This polynomial consists of all terms that have lower order than $x^{R_{\text{G-SAC},l_{d,1}}}$ in the $\hat{S}_A(x)\hat{S}_B(x)$ polynomial. This means that we can expand $\hat{S}_A(x)\hat{S}_B(x)$ as $\hat{P}_{l_{d,1}}(x) + x^{R_{\text{G-SAC},l_{d,1}}}\hat{Q}_{l_{d,1}}(x)$. The leading coefficient of $\hat{P}_{l_{d,1}}(x)$ is equal to $\sum_{i=1}^{K_d} A_{i}^{(d)}B_{i}^{(d)}$ which when summed with the earlier estimates of $\sum_{i=1}^{K_1} A_{i}^{(1)}B_{i}^{(1)},\ldots,\sum_{i=1}^{K_{d-1}} A_{i}^{(d-1)}B_{i}^{(d-1)}$, yields (in expectation) a better estimate of the $AB$ product. In the group $d\in [D]$, the layer $j \in \{2,\ldots,L_d\}$ enables the interpolation of the $\hat{P}_{l_{d,j}}(x)$ polynomial which has degree $R_{\text{G-SAC},l_{d,j}}-1 = R_{\text{G-SAC},l_{d,1}} + j - 2$. This interpolation while leading to a better estimate of $\sum_{i=1}^{K_1} A_{i}^{(1)}B_{i}^{(1)},\ldots, \sum_{i=1}^{K_d} A_{i}^{(d)}B_{i}^{(d)}$, does not provide any new estimates of the product of the remaining pairs. Therefore, its estimate improves only slightly as $j$ increases within the $d$th group.  
 
 \begin{example}[Multi-group SAC]\label{exp:gsamd}
In Fig.~\ref{fig:motiveexp}(c), we depict an example of Multi-group SAC, where $D=3$. Similar to Figs.~\ref{fig:motiveexp}(a,b), we set $K=8$ for a fair comparison. For illustrative reasons, we set other parameters as $K_d\in\{2,4,2\}$, $L_{\text{G-SAC}}=17$, $R_{\text{G-SAC},d} \in \{2,\ldots,18\}$, and $R_{\text{G-SAC}} = 19$. Compared to Example.~\ref{exp:samd}, we have $R_{\text{G-SAC},1} < 5 $ and $ 15 <  R_{\text{G-SAC}}$. In Fig.~\ref{fig:motiveexp}(c), the hatched boxes correspond to the first group of pairs: $(A_1,B_1)$ and $(A_2,B_2)$. Similarly, the grid and dotted boxes, respectively, correspond to the second and third groups of pairs. The summation of coefficients in these three groups respectively yields $\sum_{i=1}^2 A_iB_i$, $\sum_{i=3}^6 A_iB_i$, and $\sum_{i=7}^8 A_iB_i$. Compared to the arrays in Fig.~\ref{fig:motiveexp}(a,b), not only the $i$th column ($i\in [8]$) of the array in Fig.~\ref{fig:motiveexp}(c) corresponds to a different submatrix of $B$, but also the $i$th row and column do {\em not} always correspond to the coefficient of $x^{i-1}$. For example, in Fig.~\ref{fig:motiveexp}(c), the $7$th and $8$th rows correspond to the coefficients of $x^8$ and $x^9$ in $\hat{S}_A(x)$. 

Note that this choice of polynomials in generating $\hat{S}_A(x)$ and $\hat{S}_B(x)$ allows us to recover the summation of only a subset of antidiagonals in our multi-group SAC. For example, consider the grid boxes in Fig.~\ref{fig:motiveexp}(c). Although there exist four other boxes in the main antidiagonal, corresponding to the products $A_1B_7, A_2B_8,A_7B_1,$ and $A_8B_2$, the coefficient of $x^{R_{\text{G-SAC},7}-1}$ in the polynomial $\hat{S}_A(x)\hat{S}_B(x)$ is equal to the summation of the $A_iB_i$ products only for $i\in \{3,4,5,6\}$ (depicted with grid boxes on the main antidiagonal). This benefit is achieved at the cost of increasing the degree of the polynomial $\hat{S}_A(x)\hat{S}_B(x)$. While in $\hat{A}(x)\hat{B}(x)$ and $\hat{S}_A(x)\hat{S}_B(x)$ have degree 15, the degree of polynomial $\hat{S}_A(x)\hat{S}_B(x)$ is equal to $18$. This increased degree delays the exact recovery of $AB$.
\end{example}
\section{Layer-wise Successive Approximated Coding} \label{SEC:rsac}
In this section, we extend our successive approximated coding (SAC) to point-based CDC schemes including OrthoMatDot~\cite{fahim2021numerically} and Lagrange~\cite{yu2019lagrange} codes. To accomplish this, we select a specific set of evaluation points for workers to evaluate the polynomial at, different from the evaluation points used by point-based CDC schemes. Evaluating the polynomials at the elements of this specific set allows the master to approximate $AB$ starting from the evaluation of the fastest worker and continuously reducing the error as more workers report in. This continuity of approximation procedure in layer-wise SAC is in contrast to the approximation procedure of group-wise SAC, where the master realizes significant improvements in estimate quality only when each additional group of workers completes their tasks. In the following, we first detail our layer-wise SAC in terms of a general point-based CDC scheme. We then develop two specific constructions, one using OrthoMatDot~\cite{fahim2021numerically}, the other Lagrange codes~\cite{yu2019lagrange}.

%and motivates new techniques\footnote{We note that while current works on point-based CDC use either Orthonormal or Lagrange basis, the extension of point-based CDC to other well-known polynomial bases such as Newton basis can be considered for future work.}
\subsection{Layer-wise SAC for Point-based CDC}\label{sec:saomd}
We now introduce our layer-wise SAC which is applied to point-based CDC. Before doing so, we first formulate a general framework for point-based CDC. This framework unifies existing point-based methods (\hspace{1sp}\cite{fahim2021numerically,yu2019lagrange}) and motivates new techniques\footnote{We note that while current works on point-based CDC use either Orthonormal or Lagrange basis, the extension of point-based CDC to other polynomial bases can be considered for future work.}. We consider a polynomial basis $\{T_{k}(x)\}$, where the maximum degree of the first $K$ polynomials (i.e., $T_0(x),\ldots,T_{K-1}(x)$) is $K-1$. We use this basis to encode the $2K$ matrices $A_1,\ldots,A_K$ and $B_1,\ldots,B_K$. We generate encoding polynomials as $\tilde{S}_A(x)=\sum_{k=1}^K A_kT_{k-1}(x)$ and $\tilde{S}_B(x)=\sum_{k=1}^K B_kT_{k-1}(x)$. Since the degree of each $\tilde{S}_A(x)$ and $\tilde{S}_B(x)$ polynomial is (at most, and typically, is equal to) $K-1$, the degree of the decoding polynomial $\tilde{S}_A(x)\tilde{S}_B(x)$ is $2K-2$. The master assigns the multiplication of $\tilde{S}_A(x_n)$ and $\tilde{S}_B(x_n)$ to worker $n \in [N]$, where $x_n \neq x_{n'}$ for $n\neq n'$. This task equates to evaluating the decoding polynomial at a single point. In point-based CDC, each evaluation point is selected from a set of $N$ distinct real numbers, denoted $\mathcal{X}_{\text{pt-based}}$. Once workers provide the evaluation of the $\tilde{S}_A(x)\tilde{S}_B(x)$ polynomial at any $2K-1$ (distinct) points, it can be fully decoded (e.g., via the Vandermonde systems of equations). After decoding, a two-step post-decoding calculation is required to recover $AB$. In the first step, the decoding polynomial $\tilde{S}_A(x)\tilde{S}_B(x)$ is interpolated at a new set of points. At the second step, these new interpolated results are linearly combined to recover $AB$. For simplicity of applying the notation to prior point-based CDC schemes (\hspace{1sp}\cite{fahim2021numerically,yu2019lagrange}), we assume that the $\tilde{S}_A(x)\tilde{S}_B(x)$ polynomial needs to be calculated at only $K$ interpolation points $\{y_k\}_{k=1}^K$ in the first step. Then, the desired computation can be recovered via the sum $\sum_{k=1}^K \alpha_k\tilde{S}_A(y_k)\tilde{S}_B(y_k)$, where $\alpha_k$ are scalar coefficients determined by $y_k$.     

Building on the point-based CDC construction, our layer-wise SAC adds $L_{\text{L-SAC}}=2K-2$ resolution layers before exact recovery. In other words, starting from the moment that the fastest worker completes its task, the master is able (with error) to estimate $\sum_{k=1}^K \alpha_k\tilde{S}_A(y_k)\tilde{S}_B(y_k)$. The master can (in expectation) improve its estimate as more workers complete jobs. In layer-wise SAC, the master generates the same pair of polynomials $(\tilde{S}_A(x),\tilde{S}_B(x))$ as in point-based CDC schemes. However, in contrast to point-based CDC schemes which use evaluations points $x \in \mathcal{X}_{\text{pt-based}}$, in layer-wise SAC we use a different set of evaluation points, denoted $\mathcal{X}_{\text{L-SAC}}$. In particular, $\mathcal{X}_{\text{L-SAC}}=\{z_{k,i}\}_{k\in [K], i\in [n_k]}$, where $\sum_{k=1}^K n_k = N$, and for any $k\in [K]$, the $z_{k,1},\ldots,z_{k,n_k}$ are $n_k$ distinct real numbers ``$\epsilon$-close'' to $y_k$. For any $i\in [n_k]$, $y_k$ and $z_{k,i}$ are ``$\epsilon$-close'' if and only if $|y_k-z_{k,i} | \leq \epsilon$. With this definition, $z_{k,i}$ and $z_{k,j}$ are $2\epsilon$-close. Next, we explain the reason for this selection. 

%We assume that the evaluation point of worker $j_i$ is $\epsilon$-close to $y_{k_i}$, where $i\in [m]$ and $k_i \in [K]$.

To describe the decoding process, we sort workers according to their speed. We index the workers as ${j_1},\ldots,{j_N}$ so that worker $j_1$ is the fastest and worker $j_N$ is the slowest. At resolution layer $m \in [L_{\text{L-SAC}}]$, the $m$ fastest workers, corresponding to indices $j_1,\ldots,j_m$ have completed their tasks. We use $\{x_{j_1},\ldots,x_{j_m}\}$ to denote the set of evaluation points assigned to the $m$ fastest workers. Let's assume that this set includes $m_{k}$ (distinct) elements of $\{z_{k,i}\}_{i\in [n_k]}$ for each $k\in [K]$. In other words, for any $k\in [K]$, there exists (distinct) indices $j_{k,1},\ldots, j_{k,m_k}$ such that $\{x_{j_1},\ldots,x_{j_m}\} = \bigcup_{k=1}^K \{z_{k,j_{k,1}},\ldots,z_{k,j_{k,m_k}}\}$; $m_k$ must satisfy $\sum_{k=1}^K m_k = m$ and $0\leq m_k \leq n_k$. From the tasks that are completed by the $m$ fastest workers, we define 
\begin{align}\label{eq:saomd}
\tilde{C}_m = \sum_{k=1}^{K} \alpha_{k} \frac{\sum_{i=1}^{m_k}\tilde{S}_A(z_{k,j_{k,i}})\tilde{S}_B(z_{k,j_{k,i}})}{m_{k}}.
\end{align}
In limit, since polynomials are continuous   
\begin{align}\label{eq:saomd1}
\lim_{\epsilon\to 0} \tilde{C}_m = {C}_m := \sum_{k=1}^{K} \alpha_{k}\tilde{S}_A(y_{k})\tilde{S}_B(y_{k})\mathbbm{1}_{m_k > 0}, 
\end{align}
where $\mathbbm{1}_{m_k > 0}$ is an indicator function, equals 1 if $m_k > 0$, and $0$ otherwise (for detailed proof see App.~\ref{app:thm:eq}). In the following theorem, we show how to estimate $AB$ using ${C}_m$. 

\begin{theorem}\label{thm1}
 If the order of completion is uniform over all permutation, the optimal choice of $\beta$ that minimizes the expected approximation error, $\mathbb{E}\left(\|C-\beta {C}_{m}\|_F^2 \right)$, where the expectation is taken with respect to random variables $m_k$, is
\begin{align}\label{eq:LSAC:beta}
\beta^* = \frac{\left(\sum_{i=1}^K \tilde{M}_i \gamma_i\right) + \left(\sum_{i,j=1 , i< j}^K \tilde{M}_{i,j} (\gamma_{i}+\gamma_j)\right)}{\left(\sum_{i=1}^K \tilde{M}_i \gamma_i\right) + 2\left(\sum_{i,j=1, i< j}^K \tilde{M}_{i,j} \gamma_{i,j}\right)},
\end{align} 
where 
\begin{align*}
\tilde{M}_i &= \alpha_i^2 \|\tilde{S}_A(y_i)\tilde{S}_B(y_i)\|_F^2, \\
\tilde{M}_{i,j} &= \alpha_i\alpha_j \Tr\left( \left( \tilde{S}_A(y_i) \tilde{S}_B(y_i)\right)^T \left(\tilde{S}_A(y_j)\tilde{S}_B(y_j)\right)\right), \\
\gamma_i &= \frac{{N \choose m} - {N-n_i \choose m}}{{N \choose m}}, \text{and} \\
\gamma_{i,j} &= \frac{{N \choose m} - {N-n_i \choose m} - {N-n_j \choose m} + {N-n_i-n_j \choose m}}{{N \choose m}}.
\end{align*}
\end{theorem}
The proof of this theorem is provided in App.~\ref{app:thm1}. 
\begin{remark}
As in Thm.~\ref{thm0}, $\tilde{M}_i$ and $\tilde{M}_{i,j}$ are not known a priori. Furthermore, it is not always possible to approximate $\beta^*$ in a manner similar to how we did in cases~1 and~2 in Remark.~\ref{corol1}. This is because in contrast to group-wise SAC where resolution layers yield a partial sum of the $A_iB_i$ products, in layer-wise SAC (which applies to point-based CDC), ${C}_m$ is a function of the $\tilde{S}_A(y_i)\tilde{S}_B(y_i)$ rather than of the $A_iB_i$. Therefore, since the calculation of $\tilde{M}_i$ and $\tilde{M}_{i,j}$ require evaluations of the $\tilde{S}_A(y_i)\tilde{S}_B(y_i)$ products, even with prior knowledge of the distributions of $A_iB_i$, it is not apparent how to approximate $\beta^*$. However, for specific point-based polynomials where $\tilde{S}_A(y_i)\tilde{S}_B(y_i)=A_iB_i$, we can approximate $\beta^*$ when layer-wise SAC is applied to. As an example, in the next section we will show how to approximate $\beta^*$ in Lagrange layer-wise SAC.  
\end{remark}

\subsection{Examples}
We now provide examples that apply our layer-wise SAC to two specific point-based CDC schemes: OrthoMatDot~\cite{fahim2021numerically} and Lagrange~\cite{yu2019lagrange} codes.
\begin{example}[Layer-wise SAC via OrthoMatDot codes]\label{exp:saomd} 
For a fair comparison, we set $K=8$ as in Examples~\ref{exp:samd} and~\ref{exp:gsamd}. In this example, the master uses orthonormal polynomial basis $\{O_{k}(x)\}$ to generate the $7$-degree encoded polynomials via $\tilde{S}_A(x)=\sum_{k=1}^8 A_kO_k(x)$ and $\tilde{S}_B(x)=\sum_{k=1}^8 B_kO_k(x)$. For illustrative reasons, we assume that $\frac{N}{8}\in \mathbb{Z}$. The master divides $N$ workers into $K$ equally sized groups so that $n_k=N/K$ for all $k\in [8]$. It then produces the $N/8$ pairs of polynomials $(\tilde{S}_A(z_{k,1}),\tilde{S}_B(z_{k,1})),\ldots, (\tilde{S}_A(z_{k,N/8}),\tilde{S}_B(z_{k,N/8}))$, and distributes them to the $N/8$ workers of group $k \in [8]$. For all $i \in [N/8]$, $z_{k,i}$ is $\epsilon$-close to $\eta_{i}^{(8)}$. Recall from Sec.~\ref{sec:benchmark}, $\eta_{1}^{(8)},\ldots, \eta_{8}^{(8)}$ are the 8 roots of $O_8(x)$. Note that since the set $\mathcal{X}_{\text{L-SAC}}=\{z_{k,i}\}$ differs from the evaluation set used by OrthoMatDot codes ($\mathcal{X}_{\text{OMD}}=\{\eta_{n}^{(N)}\}$), our method cannot benefit from Chebyshev Vandermonde interpolation as much as OrthoMatDot codes can in the exact recovery layer. 

However, on the positive side, our Layer-wise SAC enables $L_{\text{L-SAC}}=14$ resolutions before exact recovery. Since $\tilde{S}_A(x)\tilde{S}_B(x)$ is a degree-14 polynomial, the recovery threshold is $R_{\text{L-SAC}}=15$, equal to that of Example~\ref{exp:samd} and is less than $R_{\text{G-SAC}}$ in Example~\ref{exp:gsamd}. As noted earlier, layer-wise SAC provides estimates of $AB$ that continuously improve in accuracy as workers complete tasks. This is in contrast to group-wise SAC. In Example~\ref{exp:samd}, relatively discrete steps in error reduction occur at layers $l =1$ and $l=9$. In Example~\ref{exp:gsamd}, such steps occur at layers $l=1, 7,$ and $17$. 
\end{example}

\begin{example}[Layer-wise SAC via Lagrange codes] \label{exp:slag}
In contrast to the previous example that uses an orthonormal basis, we now extend layer-wise SAC to Lagrange codes which use the Lagrange basis $\{L_k(x)\}$. As before, let $K=8$ and set $n_k=N/8$ for all $k \in [8]$. In this example, our method generates encoding polynomials that are similar to those of Lagrange codes: $\tilde{S}_A(x)=\sum_{k=1}^8 A_kL_k(x)$ and $\tilde{S}_B(x)=\sum_{k=1}^8 B_kL_k(x)$. However, in layer-wise SAC, these polynomials are evaluated at $\mathcal{X}_{\text{L-SAC}}=\{y_{k,i}\}$, where $y_{k,i}$ is $\epsilon-$close to $y_k$ for $i \in [N/8]$ and $k \in [8]$. Similar to Example~\ref{exp:saomd}, when layer-wise SAC method applies to Lagrange codes, it provides its first estimate when the first worker completes its task (note $R_{\text{L-SAC},1}=1$). Layer-wise SAC then continuously improves its estimate as $l$ increases from 2 to 14. Layer-wise SAC recovers the exact $AB$ product once $R_{\text{L-SAC}}=15$ workers complete their tasks.    

There are some differences between Examples~\ref{exp:saomd} and~\ref{exp:slag}. In addition to the use of different polynomial bases and different evaluation points, in Example~\ref{exp:slag} the $AB$ product can be recovered by simply computing the sum $\sum_{k=1}^K \tilde{S}_A(y_k)\tilde{S}_B(y_k)$. I.e., in Example~\ref{exp:slag} the $k$th scalar coefficient equals $\alpha_k=1$. However, in Example~\ref{exp:saomd}, $\tilde{S}_A(y_k)\tilde{S}_B(y_k)$ should first be scaled by $\frac{2}{K}$ and then summed. Furthermore, in contrast to Example~\ref{exp:saomd}, the evaluation of $\tilde{S}_A(y_k)\tilde{S}_B(y_k)$ in Example~\ref{exp:slag} results in the $A_kB_k$ matrix product. Therefore, with prior knowledge of the distributions of $A_kB_k$, we can approximate $\beta^*$ in Lagrange layer-wise SAC. Similar to Remark~\ref{corol1}, we consider the following two cases to approximate $\beta^*$ in (\ref{eq:LSAC:beta}).
\begin{itemize}
\item \textit{Case 1:} If for any $i\in [K]$, $n_i=N/K$, and $\sum_{i,j=1,i<j}^K \tilde{M}_{i,j} \ll \sum_{i=1}^K \tilde{M}_{i}$, then (\ref{eq:LSAC:beta}) is simplified to $\beta^*=1$. In Example~\ref{exp:slag}, since $\tilde{M}_{i}=\|A_iB_i\|_F^2$ and $\tilde{M}_{i,j}=\Tr \left(\left(A_iB_i\right)^T\left(A_jB_j\right) \right)$, condition $\sum_{i,j=1,i<j}^K \tilde{M}_{i,j} \ll \sum_{i=1}^K \tilde{M}_{i}$ holds (in expectation) if the entries of $A_i$ and $B_i$ are i.i.d and have zero mean and high variance.
\item \textit{Case 2:} If for any $i\in [K]$, $n_i=N/K$, and $\sum_{i,j=1,i<j}^K \tilde{M}_{i,j} \gg \sum_i \tilde{M}_{i=1}^K$, then (\ref{eq:LSAC:beta}) is simplified to
\begin{align}\label{eq:optbetaExp}
\beta^*\approx \frac{\gamma_{i} + \gamma_{j} }{2\gamma_{i,j}} = \frac{\left({N \choose m} - 2{N-N/K \choose m} + {N-2N/K \choose m}\right)}{\left({N \choose m} - {N-N/K \choose m}\right)}.
\end{align}
In Example~\ref{exp:slag}, the condition $\sum_{i,j=1,i<j}^K \tilde{M}_{i,j} \gg \sum_{i=1}^K \tilde{M}_{i}$ can be obtained (in expectation) when the entries of $A_i$ and $B_i$ are independent and have mean zero, and $A_i$ (and $B_i$) are strongly and positively correlated. Two matrices $A_i$ and $A_j$ are said to be correlated if any $(e,k)$th entries in $A_i$ and $A_j$, i.e., $(A_i)_{e,k}$ and $(A_j)_{e,k}$, are correlated.  
\end{itemize}
%we reevaluate the encoding polynomials at points in $\mathcal{X}_{\text{L-SAC}}=\{y_{k,i}\}$, while in Example.~\ref{exp:saomd} the evaluation points are $\{z_{k,i}\}$. The evaluation points for layer-wise SAC design of Lagrange codes are $\epsilon$-close to an arbitrary set of distinct reals ($\{y_k\}_{k=1}^8$). In contrast, in the layer-wise SAC design of OrthoMatDot codes, the evaluation points are specifically selected to be $\epsilon$-close to $\{\eta_{k}^{(8)}\}_{k=1}^8$, the roots of the $O_8(x)$ polynomial. In Sec.~\ref{sec:simulation}, we provide simulations that demonstrate how the choice of $\mathcal{X}_{\text{G-SAC}}$ and $\mathcal{X}_{\text{L-SAC}}$ affects the numerical stability of the results.

%There are some differences between Examples~\ref{exp:saomd} and~\ref{exp:slag}. Besides using different encoding polynomials, in Example~\ref{exp:slag} we evaluate the encoding polynomials at points in $\mathcal{X}_{\text{L-SAC}}=\{y_{k,i}\}$, while in Example.~\ref{exp:saomd} the evaluation points are $\{z_{k,i}\}$. The evaluation points for layer-wise SAC design of Lagrange codes are $\epsilon$-close to an arbitrary set of distinct reals ($\{y_k\}_{k=1}^8$). In contrast, in the layer-wise SAC design of OrthoMatDot codes, the evaluation points are specifically selected to be $\epsilon$-close to $\{\eta_{k}^{(8)}\}_{k=1}^8$, the roots of the $O_8(x)$ polynomial. In Sec.~\ref{sec:simulation}, we provide simulations that demonstrate how the choice of $\mathcal{X}_{\text{G-SAC}}$ and $\mathcal{X}_{\text{L-SAC}}$ affects the numerical stability of the results.
 \end{example}

\section{Simulation Results} \label{sec:simulation}
In this section, we experimentally compare our SAC methods and benchmark them against the state-of-the-art $\epsilon$-approximate MatDot codes~\cite{9509407}. We conduct experiments in Python using double-precision floating-point numbers with machine epsilon approximately equal to $2.22\times 10^{-16}$. We simulate the task of computing a matrix product when the task is distributed across $N=24$ workers. Throughout this section, the task is to multiply a $100\times 8000$ matrix $A$ with an $8000 \times 100$ matrix $B$. Unless otherwise specified, the entries of $A$ and $B$ are real numbers selected independently from zero-mean and unit variance normal distributions. We next detail how we simulate the distributed implementation of the matrix product.

In each experiment, we first implement the encoder of the different CDC schemes. For each CDC scheme, we generate $N$ pairs of evaluated encoding polynomials and multiply them to produce $N$ different evaluations of the decoding polynomial. We then {\em uniformly} shuffle these $N$ evaluated decoding polynomials, where the $m$th element will correspond to the decoding polynomial computed by the $m$th fastest worker. The decoder that is implemented based on the first $m\in [N]$ evaluations outputs a matrix $\tilde{C}_m \in \mathbb{R}^{100\times 100}$, an estimate of $AB$. For each coding scheme, we report the relative error between $AB$ and $\tilde{C}_m$ starting with the first $m\in [N]$ such that the scheme can provide an estimate. Due to the randomness in the permutation of $(A_i,B_i)$ pairs and the randomness in the order of workers' completion, we repeat each experiment $100$ times and report average relative errors.

In Sec.~\ref{sec:comp_vs_app}, we characterize two sources of error, approximation and computation errors, both of which affect the relative error. We provide numerical experiments that compare group-wise and layer-wise SAC in terms of these sources of error. To benchmark our SAC methods against $\epsilon$-approximate MatDot codes, in Sec.~\ref{sec:sim:tradeoff}, we numerically explore the tradeoff between relative error and approximation threshold. Based on Theorems~\ref{thm0} and~\ref{thm1}, we optimally set the parameters of SAC showing that SAC outperforms $\epsilon$-approximate MatDot codes in at least two ways. First, SAC lowers the approximation threshold of $\epsilon$-approximate MatDot codes. Second, SAC reduces the relative error given the same approximation threshold used by $\epsilon$-approximate MatDot codes.

%we numerically characterize a  fundamental tradeoff between approximate recovery and the quality of approximation. The objective of this tradeoff is to  minimize the number of workers required to return their computed results (denoted by the approximation threshold) to estimate the desired computation within a certain degree of error. Using this tradeoff, we (analytically and theoretically) prove that each of our proposed SAC methods has their own benefits. 

\subsection{Sources of Errors} \label{sec:comp_vs_app}
We now consider two distinct sources of error: approximation and computation. Approximation error refers to the difference between the original $AB$ product and its {\em best} approximation that can be derived analytically when only a subset of workers complete their tasks. Recall from Sec.~\ref{sec:SAC} and~\ref{SEC:rsac} that we used ${C}_m$ to denote the best (analytical) approximate of $C=AB$ which can be recovered when the $m$ fastest workers report in, and we term $\|C-{C}_m\|_F^2$ the approximation error. This error results from the truncation in the calculation contributed to by only a subset of workers and neglects the error caused by numerical computation and estimation. In contrast, computation error refers to the difference between the $C_m$ and its estimate that is computed in reality. There are two major sources of computation error. One is the numerical error in the encoding and decoding computations. The other one is the error of estimating the decoding polynomial which is used for approximation recovery (e.g., the error of estimating $\hat{P}_l(x)$ by $\hat{S}_A(x)\hat{S}_B(x)$ in two-group SAC). As before, $\tilde{C}_m$ is used to denote the estimated computation of $C_m$. In other words, the approximate $C_m$ is an exact version of $\tilde{C}_m$. We term $\|{C}_m-\tilde{C}_m\|_F^2$ the computation error. 

While approximation and computation errors have been (separately) analyzed in prior work~\cite{fahim2021numerically,ramamoorthy2021numerically, 9509407}, in this paper, we introduce the notion of {\em total} error that accounts for both approximation and computation errors. For simplicity of notation, we denote the total error by {\em error} in this paper. Applying the triangle inequality to the Frobenius norm, we can bound the total error in terms of the approximation and computation errors as
%\begin{align}\label{eq:boundEr}
%\underbrace{\sqrt{\|C - \tilde{C}_m\|_F^2}}_{\text{Error}} \leq \underbrace{\sqrt{\|C - {C}_m\|_F^2}}_{\text{Approximation error}} + \underbrace{\sqrt{\|{C}_m - \tilde{C}_m\|_F^2}}_{\text{Computation error}} .
%\end{align}  
\begin{align}\label{eq:boundEr}
\sqrt{\vphantom{\|C - \tilde{C}_m\|_F^2}\smash{
  \underbrace{\|C - \tilde{C}_m\|_F^2}_{\text{Error}}
  }} \leq \sqrt{\vphantom{\|C - {C}_m\|_F^2}\smash{
  \underbrace{\|C - {C}_m\|_F^2}_{\text{Approximation error}}
  }} + \sqrt{\vphantom{\|C - \tilde{C}_m\|_F^2}\smash{
  \underbrace{\|C_m - \tilde{C}_m\|_F^2}_{\text{Computation error}}
  }} \\ \nonumber
\end{align}

\begin{figure*}[!htbp]%[h] %!t
\centering  
       	\subfloat[G-SAC, $\epsilon=0.45$ for $\mathcal{X}_{\text{equal}}$, $\epsilon=0.15$ for $\mathcal{X}_{\text{complex}}$]{\includegraphics[width=7.5cm]{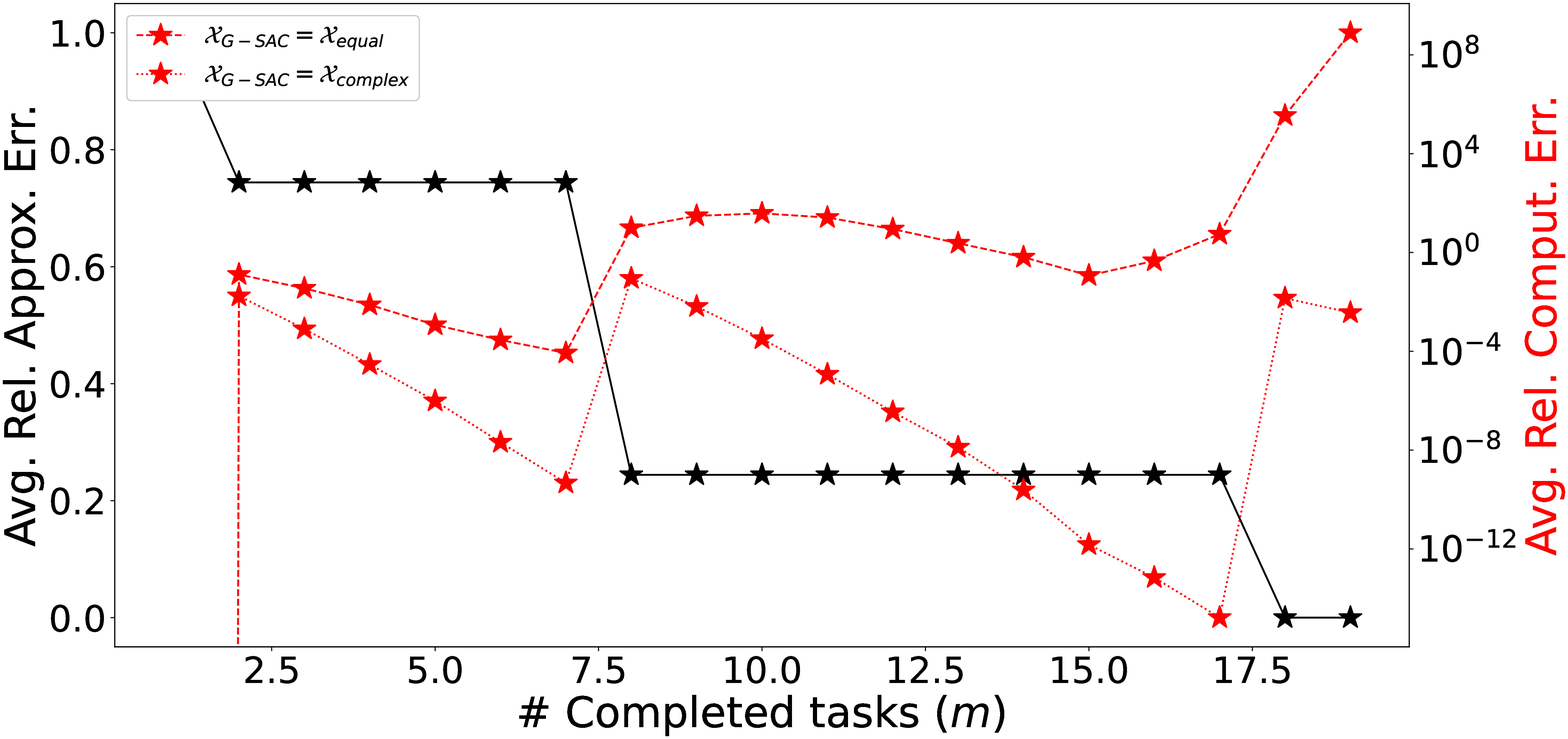}        
       \label{FIG:tradeoff2} }
       	\quad 
       	 \subfloat[L-SAC, $\epsilon=0.0125$ for $\mathcal{X}_{\text{L-SAC}}$]{\includegraphics[width=7.5cm]{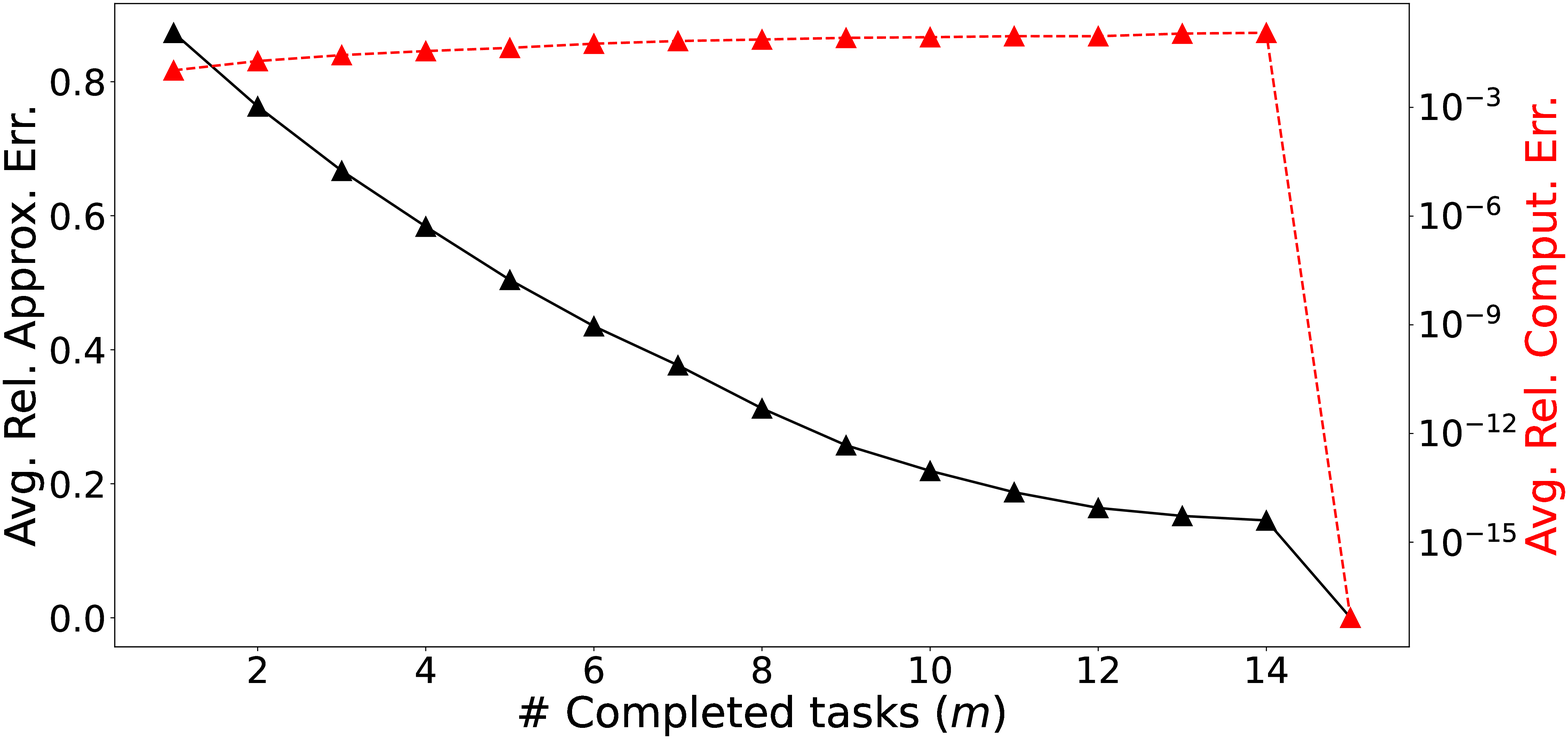}        
       \label{FIG:tradeoff4} }
       	\quad 
       		\subfloat[G-SAC, $\#$ Completed tasks$=8$]{\includegraphics[width=7.5cm]{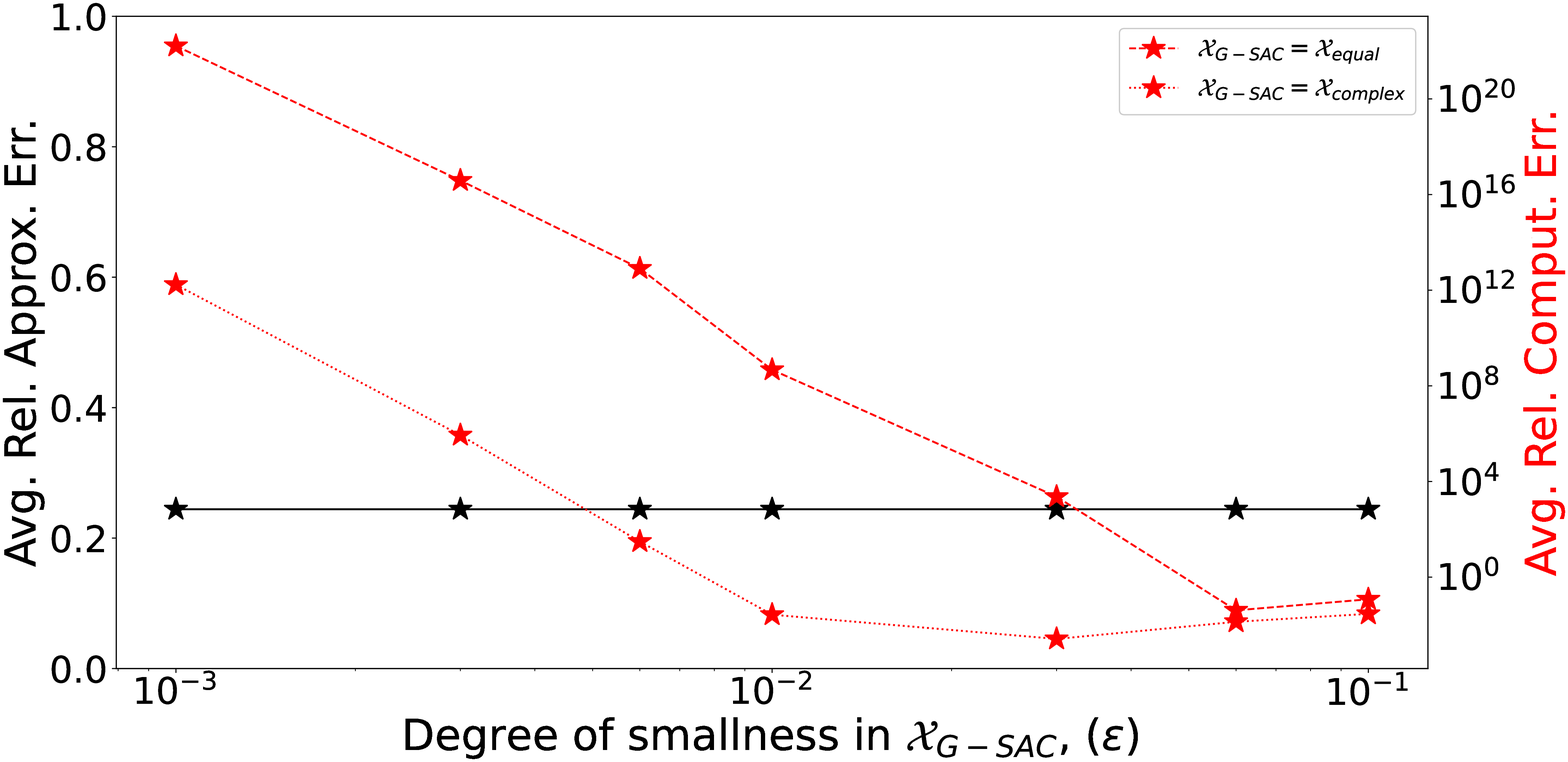}        
       \label{FIG:tradeoff1} }
       	\quad   	
       	 \subfloat[L-SAC, $\#$ Completed tasks$=8$]{\includegraphics[width=7.5cm]{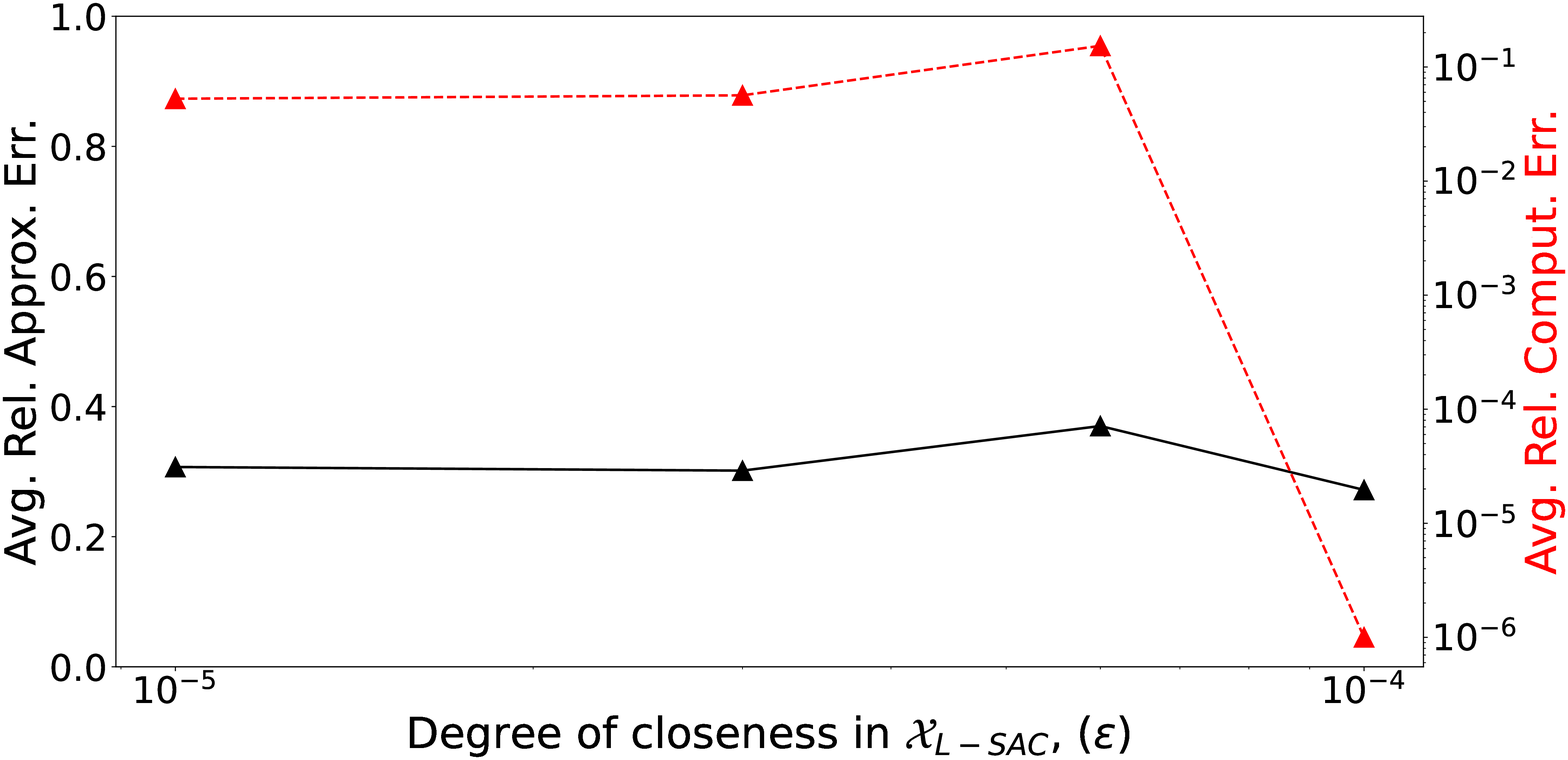}        
       \label{FIG:tradeoff3} }
       	\quad    
       %	\subfloat[L-SAC, $\mathcal{X}_{\text{ortho}}$, No. completed tasks$=8$]{\includegraphics[width=8cm]{./image/tradeoff_ErrVsEps_LSAC_normalAB_No8.eps}        
       %\label{FIG:tradeoffcont} }
       	%\quad      	
	%\subfloat[L-SAC, $\mathcal{X}_{\text{ortho}}$, $\epsilon=0.012$]{\includegraphics[width=8cm]{./image/tradeoff_ErrVsRT_LSAC_normalAB_logscale.eps}        
     %  \label{FIG:tradeoffcont} }
      % 	\quad
       \caption{A dual $y$-axis (average relative approximation error, average relative computation error) vs. number of completed tasks in (a,b); vs. $\epsilon$ of $\mathcal{X}_{\text{G-SAC}}$ in (c); and vs. $\epsilon$ of $\mathcal{X}_{\text{L-SAC}}$ in (d). In (a,b), $\epsilon=0.45$ if $\mathcal{X}_{\text{G-SAC}}=\mathcal{X}_{\text{equal}}$, and $\epsilon=0.15$ if $\mathcal{X}_{\text{G-SAC}}=\mathcal{X}_{\text{complex}}$. In (c,d), number of workers is equal to $8$.} \label{FIG:tradeoff}
\end{figure*}

In the sequel, we define the {\em relative} error, where we normalize each error by $\|C\|_F^2$. Note that Figs~\ref{FIG:tradeoff}(a)-(d) are dual $y$-axis plots. The left $y$-axis corresponds to the average relative approximation error and corresponds to the black solid curves. The right $y$-axis corresponds to the average relative computation error and corresponds to the red dashed or dotted curves. In Figs.~\ref{FIG:tradeoff2} and~\ref{FIG:tradeoff4}, we plot these two quantities versus the number of completed tasks for group-wise and layer-wise SAC. We use a group-wise SAC (G-SAC) that consists of $G=3$ groups. As in Example~\ref{exp:gsamd}, we set $K=8$, and for each group we respectively set $K_1=2, K_2=4, $ and $K_3 = 2$. In G-SAC, we consider two different choices of evaluation points. First, we select evaluation points to be equidistant on the real line, in particular the set $\mathcal{X}_{\text{equal}} = \{\frac{\epsilon n}{N}\}_{n=1}^{N} $, where $N=24$ and $\epsilon$ is small. In Fig.~\ref{FIG:tradeoff2}, we fix $\epsilon$ to be equal to $0.45$, and in Fig.~\ref{FIG:tradeoff1} we present results for a variety of choices for $\epsilon$, $\epsilon \in \{10^{-3}, 3\times 10^{-3}, 6\times 10^{-3}, 10^{-2}, 3\times 10^{-2},6\times 10^{-2}, 10^{-1} \}$. Second, we select the $N$ evaluation points to be complex and at equal-magnitude, $\mathcal{X}_{\text{complex}} = \{\epsilon e^{i2\pi n / 24}\}_{n=1}^{24} $, where $\epsilon$ equals $0.15$ in Fig.~\ref{FIG:tradeoff2} and varies, $\epsilon \in \{10^{-3}, 3\times 10^{-3}, 6\times 10^{-3}, 10^{-2}, 3\times 10^{-2},6\times 10^{-2}, 10^{-1}\}$ in Fig.~\ref{FIG:tradeoff1}. The red dashed curves plot results for $\mathcal{X}_{\text{G-SAC}}=\mathcal{X}_{\text{equal}}$, and the red dotted curves for $\mathcal{X}_{\text{G-SAC}}=\mathcal{X}_{\text{complex}}$. For layer-wise SAC (L-SAC), we use OrthoMatDot codes~\cite{fahim2021numerically} using a similar setting to Example~\ref{exp:saomd}. We use $K=8$ for a fair comparison and the evaluation set equals $\mathcal{X}_{\text{L-SAC}}=\{z_{k,i}\}_{k \in [8], i \in [3]}$, where the $z_{k,i}$ are $\epsilon$-close to $\eta_{i}^{(8)}$ (the roots of $O_8(x)$). In Fig.~\ref{FIG:tradeoff4}, we fix $\epsilon$ to be equal to $0.0125$, and in Fig.~\ref{FIG:tradeoff3} we vary $\epsilon \in \{10^{-5}, 3\times 10^{-5}, 6\times 10^{-5}, 10^{-4}\}$.

Considering the black solid curves in Figs.~\ref{FIG:tradeoff2} and~\ref{FIG:tradeoff4}, we observe that the average relative approximation error is non-increasing in the number of workers ($m$) that have completed their tasks. This meets our expectations and can be explained via the concept of underfitting. When $m$ is small, a smaller number of workers have completed their tasks and thus a decoder of lower complexity tends to underfit the computations that have been completed thus far. For example, in the case of polynomial interpolation, a lower degree polynomial needs to be interpolated when $m$ is small. As $m$ increases, more workers complete their tasks and a more complex decoder must be used to recover the desired $AB$ product. This monotonically non-increasing change in relative approximation error experiences three distinct drops in error in Fig.~\ref{FIG:tradeoff2}. These drops occur when groups of layers complete, i.e., when $m$ equals $R_{\text{G-SAC},1}=2, R_{\text{G-SAC},7}=8, $ or $R_{\text{G-SAC},17}=18$. At these values of $m$, the estimate $\tilde{C}_m$ is formed by adding $\sum_{j=1}^{2} A_{i_j}B_{i_j}$, $\sum_{j=3}^{6} A_{i_j}B_{i_j}$, and $\sum_{j=7}^{8} A_{i_j}B_{i_j}$. On the other hand, Fig.~\ref{FIG:tradeoff4} indicates a smooth decrease of the average relative approximation error for L-SAC as $m$ increases. At $m=15$, exact recovery is possible and an average relative approximation error of zero is obtained.

The change in error as $m$ increases, while being monotonically non-increasing for relative approximation error, is not monotonic for relative computation error. In Fig.~\ref{FIG:tradeoff2}, as the red dashed and dotted lines show, the relative computation errors increase at specific values of $m \in \{2,8,18\}$ and decrease in $m$ for other values. To derive an explanation we must consider two opposing trends. On the one hand, we note that an increase in $m$ makes the decoding procedure more complex. More complex decoding is more subject to numerical instability~\cite{fahim2021numerically}. A larger computation error is the result. On the other hand, as $m$ increases, the decoder used by approximate recovery estimates the decoder of exact recovery more precisely. Therefore, the estimated computation $\tilde{C}_m$ should converge to $C_m$ as $m$ increases, lowering the average relative computation error. 

We now consider these two opposing trends as $m$ increases. In Fig.~\ref{FIG:tradeoff2}, the average relative approximation error of G-SAC increases only when $m \in \{2,8,18\}$. At these values of $m$, an additional coefficient of the decoding polynomial (corresponding to $\sum_{j=1}^{2} A_{i_j}B_{i_j}$, $\sum_{j=3}^{6} A_{i_j}B_{i_j}$, or $\sum_{j=7}^{8} A_{i_j}B_{i_j}$) contributes to the $AB$ recovery. When a new coefficient of a decoding polynomial is added to $\tilde{C}_m$, numerical stability issues that affect the computation error become more significant. But, when $m$ increases for other values ($m\neq 2,8,$ or 18), then $\tilde{C}_m$ gets close to $C$ and we observe a decrease in average relative computation error. In Fig.~\ref{FIG:tradeoff4}, the red dashed line shows that the average relative computation error of L-SAC increases slightly as $m$ increases up to $m=15$. This is because $m=\sum_{k=1}^K m_k$ and as $m$ increases from 1 to 14, the more likely $m_k$ random variables become non-zero. Recalling the definitions of $\tilde{C}_m$ and ${C}_m$ from (\ref{eq:saomd}) and (\ref{eq:saomd1}),  $m_k\neq 0$ for more indices $k\in [8]$ means that more non-zero terms are added to form both $\tilde{C}_m$ and ${C}_m$. The terms that are newly added to $\tilde{C}_m$ differ from those added to ${C}_m$. This increases the average relative computation error. However, when $m=15$, L-SAC can recover $C$ almost exactly, and thus a negligible average relative computation error ($\approx 10^{-17}$) is observed.

One factor that affects the relative computation error is the choice of evaluation sets, $\mathcal{X}_{\text{G-SAC}}$ and $\mathcal{X}_{\text{L-SAC}}$. Figures~\ref{FIG:tradeoff2} and~\ref{FIG:tradeoff1} show that the choice $\mathcal{X}_{\text{G-SAC}}=\mathcal{X}_{\text{complex}}$ outperforms the choice $\mathcal{X}_{\text{G-SAC}}=\mathcal{X}_{\text{equal}}$, as the former (the red dotted curves) have a lower average relative computation error than the latter (the red dashed curves). We can understand this as resulting from the condition number of the Vandermonde matrix in the G-SAC decoder. When evaluation points are selected from $\mathcal{X}_{\text{equal}}$, the G-SAC decoder solves a system of linear equations that involves a real Vandermonde matrix. However, when sampling using $\mathcal{X}_{\text{complex}}$, the entries of the Vandermonde matrix are complex entries of equal magnitude, i.e., located on a circle in $\mathbb{C}$. As is shown in~\cite{ramamoorthy2021numerically}, the condition number of a real Vandermonde matrix exponentially grows in $m$, while when the entries are chosen from $\mathcal{X}_{\text{complex}}$ the condition number of the Vandermonde matrix  grows only polynomially in $m$. Ill-conditioning leads to numerical problems which contribute to the relative computation error. This explains the superiority of $\mathcal{X}_{\text{complex}}$ over $\mathcal{X}_{\text{equal}}$ in terms of relative computation error. That said, since each complex multiplication is equivalent to four real multiplications, the use of complex evaluation points from $\mathcal{X}_{\text{complex}}$ increases the computation load of each worker by a factor of four when compared to the use of real evaluation points in $\mathcal{X}_{\text{equal}}$. Thus, despite lower relative computation error, a downside to the use of $\mathcal{X}_{\text{complex}}$ is the increased computation.

Another way to lower the relative computation error of G-SAC is to set its evaluation points to be sufficiently small. In Fig.~\ref{FIG:tradeoff1}, we fix $m=8$ and show that the average relative computation error of G-SAC is minimized by setting $\epsilon=3\times 10^{-2}$ in $\mathcal{X}_{\text{complex}}$ and to $\epsilon=6\times 10^{-2}$ in $\mathcal{X}_{\text{equal}}$. Similarly, in Fig,~\ref{FIG:tradeoff3}, we show that the average relative computation error of L-SAC is minimized by setting $\epsilon=10^{-4}$ in $\mathcal{X}_{\text{L-SAC}}$. We now make two comments. First, note that as both Figs.~\ref{FIG:tradeoff1} and~\ref{FIG:tradeoff3} indicate, setting $\epsilon$ lower than some threshold increases the average relative computation error. This is due to the finite precision of simulations. Second, in contrast to computation error, the relative approximation error is independent of $\epsilon$. The black solid curves in Figs.~\ref{FIG:tradeoff1} and~\ref{FIG:tradeoff3} show that the average relative approximation errors of both G-SAC and L-SAC are approximately equal to $0.3$ when we fix $m=8$ and vary $\epsilon$.

\subsection{Fundamental Tradeoffs}\label{sec:sim:tradeoff}
In contrast to the previous section, where we evaluated relative approximation error and relative computation error separately, in this section we jointly consider these two sources of error. We label curves for two group-wise SAC (G-SAC), for two layer-wise SAC (L-SAC), and for $\epsilon$-approximate MatDot codes~\cite{9509407}  using, respectively, star, triangular, and square marks. To ensure a fair comparison, we set $K=8$ for all curves. For the G-SAC implementations, we use two-group SAC twice, one with $K_1=5$ and the other with $K_1=8$. In both, we select $N$ evaluation points from $\mathcal{X}_{\text{complex}}=\{0.1 e^{i2\pi n / N}\}_{n=1}^{N}$, where $N=24$. For the two L-SAC implementations, we use OrthoMatDot~\cite{fahim2021numerically} and Lagrange~\cite{yu2019lagrange} codes with settings similar to those used in Examples~\ref{exp:saomd} and~\ref{exp:slag}. In particular, for OrthoMatDot L-SAC, the evaluation set $\mathcal{X}_{\text{L-SAC}}$ equals $\{z_{k,i}\}_{k\in[8], i\in[3]}$, where the $z_{k,i}$ are $\epsilon$-close to $\eta_k^{8}$ and $\epsilon=6.25\times 10^{-3}$. For Lagrange L-SAC, $\mathcal{X}_{\text{L-SAC}}$ equals $\{y_{k,i}\}_{k\in[8], i\in[3]}$, where the $y_{k,i}$ are $\epsilon$-close to $y_k$, $\epsilon=3.33\times 10^{-2}$, and $y_k =k$ for any $k\in [8]$.

\begin{figure}[!htbp]%[!t]
\centering 
	\subfloat[$\lambda = 0$]{\includegraphics[width=7.5cm]{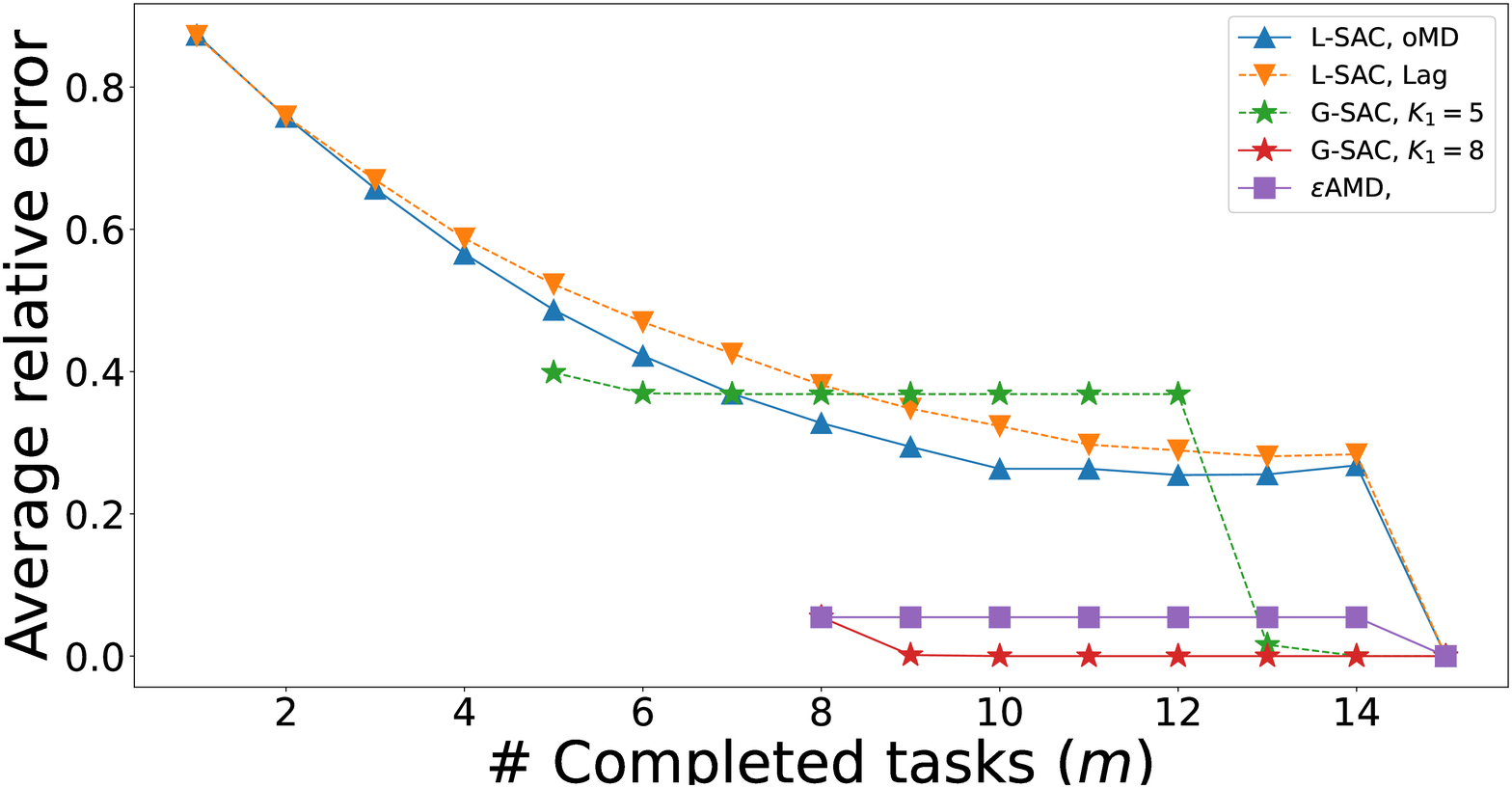}        
       \label{FIG:variable1} }
       	\quad
	\subfloat[$m = 8$]{\includegraphics[width=7.5cm]{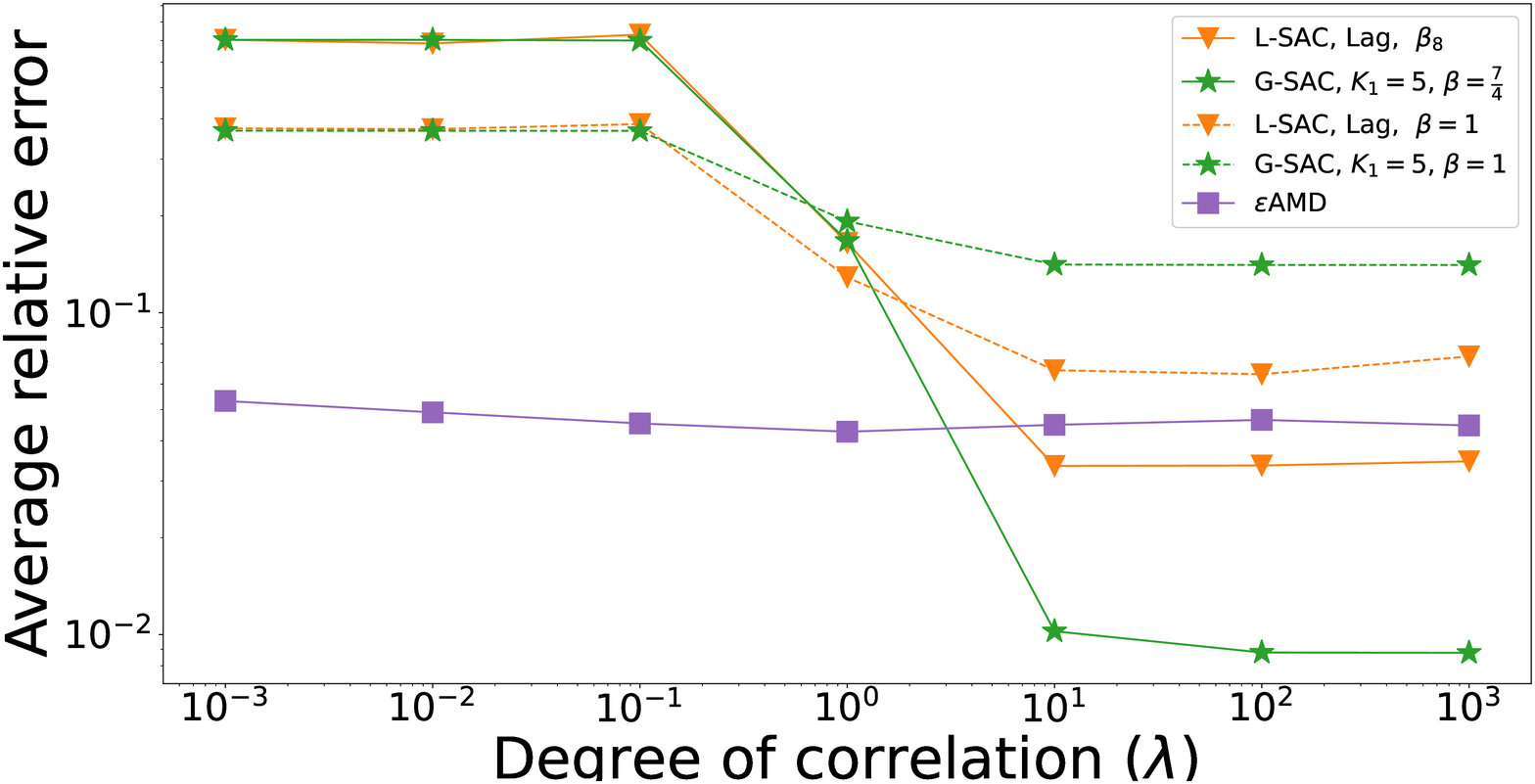}          \label{FIG:variable2} }  	\quad
     %\subfloat[]{\includegraphics[width=8cm]{./image/nunifAB_N24_8GSAC_55442211.eps}        \label{FIG:3} } 	\quad
     % \subfloat[]{\includegraphics[width=8cm]{./image/tradeoff_continious.eps}         \label{FIG:4} }	\quad
       \caption{Average relative error vs. (a) the number of completed tasks ($m$), where $\lambda=0$; vs. (b) the degree of correlation ($\lambda$), where $m=8$.} 
\end{figure}

\textbf{Relative error vs. approximation threshold:} We first plot average relative error vs. the number of completed tasks ($m$) in Fig.~\ref{FIG:variable1}. Note that the $m$ that varies from 1 to 14 is a variable representing the approximation threshold. The $m=15$ is the recovery threshold. Figure~\ref{FIG:variable1} shows that while all CDC schemes are able to recover $AB$ with (almost) zero error when $m=15$, they achieve different tradeoffs between average relative error and approximation threshold (when $m<15$). As Fig.~\ref{FIG:variable1} shows the $\epsilon$-approximate MatDot code provides an estimate of $AB$, only once when $m=8$. It then needs to wait for exact recovery (until $m=15$) to improve on this estimate. However, the other four SAC methods can improve on this tradeoff even when $m < 15$. 

The G-SAC method (the red solid curve) must wait until $m=8$ to provide its first estimate; This is similar to $\epsilon$-approximate MatDot codes. However, in contrast to $\epsilon$-approximate MatDot codes, the G-SAC method improves on this estimate as $m$ increases by achieving a close to zero average relative error before exact recovery. The G-SAC method (the green dashed curve) can provide an earlier estimate (when $m=5$) compared to the previous two methods. It then gradually improves on this estimate when $m$ increases up to $m=13$. However, when $m<13$, the estimate that this G-SAC method provides has a higher average relative error, compared to the previous two methods. When $m=13$, the average relative error of the second G-SAC method displays a significant drop such that it outperforms $\epsilon$-approximate MatDot codes. Indeed, this G-SAC method can achieve an average relative error close to zero when $m=14$; similar to the first G-SAC method. 

The two L-SAC methods are able to produce estimates of $AB$ since $m=1$, and continuously improve on average relative error as $m$ increases above 1. However, as shown in Fig.~\ref{FIG:variable1}, only when $7 \leq  m < 13$ can the OrthoMatDot L-SAC method provide a lower average relative error than can the second G-SAC method. This range of $m$ is even smaller for the Lagrange L-SAC method. This method only outperforms the second G-SAC method in terms of average relative error when $9 \leq m < 13$. Figure~\ref{FIG:variable1} shows that despite providing earlier estimates, L-SAC methods do not achieve lower average relative error when compared to the first G-SAC and $\epsilon$-approximate MatDot codes. That said, we next describe situations in which Lagrange L-SAC methods can achieve lower average relative error than $\epsilon$-approximate MatDot code even when $m\geq 8$.

\textbf{Effect of correlation:} We note that in all previous figure, the entries of $A$ and $B$ are assumed to be independently selected from the zero-mean and unit variance normal distribution. This populates $A$ and $B$ matrices with both positive and negative entries, and the $A_iB_i$ matrices are uncorrelated for all $i\in [8]$. With respect to Remark~\ref{corol1}, we define two matrices $A_iB_i$ and $A_jB_j$ to be uncorrelated if 
\begin{align}\label{eq:correlatedcond}
\Tr\left(\mathbb{E}\left((A_iB_i)^T(A_jB_j)\right)\right)=\Tr\left(\mathbb{E}\left(A_iB_i\right)^T\mathbb{E}\left(A_jB_j\right)\right)
\end{align}
and to be correlated otherwise. This condition is obtained directly from the assumption that any $(A_iB_i)_{e,k}$ and $(A_jB_j)_{e,k}$ entries in $A_iB_i$ and $A_jB_j$ are uncorrelated. In other words,~(\ref{eq:correlatedcond}) is obtained based on the assumption that $\mathbb{E}((A_iB_i)_{e,k}(A_jB_j)_{e,k})=\mathbb{E}((A_iB_i)_{e,k})\mathbb{E}((A_jB_j)_{e,k})$. We next consider a situation where $\{A_iB_i\}$ are correlated. We construct the $A_i$ and $B_i$ matrices as $A_i= \lambda A^{(0)} + A_i^{(1)}$ and $B_i= \lambda B^{(0)} + B_i^{(1)}$, where $\lambda$ is the degree of correlation. In this construction, $A^{(0)}$ and $B^{(0)}$ are latent random matrices common to all the $A_i$ and $B_i$. The $A_i^{(1)}$ and $B_i^{(1)}$ are random matrices that are specific to the $i$th matrices $A_i$ and $B_i$. As before, the entries of the $A^{(0)}$, $A_i^{(1)}$, $B^{(0)}$, and $B_i^{(1)}$ matrices are selected independently from the zero-mean and unit-variance normal distribution for all $i \in [8]$. The correlation degree $\lambda$ plays a key role. The correlation between $A_iB_i$ and $A_jB_j$ when $i\neq j$ is
\begin{align}\label{eq:corr1}
\mathbb{E}\left(\Tr\left((A_iB_i)^T(A_jB_j)\right)\right) =\lambda^4\mathbb{E}\left(\|A^{(0)}B^{(0)}\|_F^2\right),
\end{align}
while when $i=j$ is
\begin{align}\label{eq:auto1}
&\mathbb{E}\left(\Tr\left((A_iB_i)^T(A_iB_i)\right)\right) = \mathbb{E}\left(\|A_iB_i\|_F^2\right) \nonumber \\ &=\lambda^4\mathbb{E}\left(\|A^{(0)}B^{(0)}\|_F^2\right)  + \lambda^2\mathbb{E}\left(\|A^{(0)}B_i^{(1)}\|_F^2\right) + \nonumber\\ &+ \lambda^2\mathbb{E}\left(\|A_i^{(1)}B^{(0)}\|_F^2\right) + \mathbb{E}\left(\|A_i^{(1)}B_i^{(1)}\|_F^2\right).
\end{align} 
% \begin{align}\label{eq:auto1}
%&\mathbb{E}\left(\Tr\left((A_iB_i)^T(A_iB_i)\right)\right) = \mathbb{E}\left(\|A_iB_i\|_F^2\right) \nonumber \\ &=\lambda^4\mathbb{E}\left(\|A^{(0)}B^{(0)}\|_F^2\right)  + \lambda^2\mathbb{E}\left(\|A^{(0)}B_i^{(1)}\|_F^2\right) + \nonumber\\ &+ \lambda^2\mathbb{E}\left(\|A_i^{(1)}B^{(0)}\|_F^2\right) + \mathbb{E}\left(\|A_i^{(1)}B_i^{(1)}\|_F^2\right).
%\end{align} 
% two col

We now consider two cases. First, when $\lambda$ is close to zero. In this case,~(\ref{eq:corr1}) is almost zero while~(\ref{eq:auto1}) equals $\mathbb{E}\left(\|A_i^{(1)}B_i^{(1)}\|_F^2\right)$, which is nonzero. Second, when $\lambda \gg 1$, both~(\ref{eq:corr1}) and~(\ref{eq:auto1}) are almost equal to $\lambda^4\mathbb{E}\left(\|A^{(0)}B^{(0)}\|_F^2\right)$. These two cases can be viewed in analogy with the two cases considered in Remark~\ref{corol1} and Example~\ref{exp:slag}. Accordingly, when $\lambda\approx 0$, the approximation error of G-SAC can be minimized if we set $\beta=1$ in (\ref{eq:thm0beta}). Similarly, the L-SAC method that applies to Lagrange codes achieves the minimum approximation error if $\beta$ in (\ref{eq:LSAC:beta}) is set to 1. When $\lambda \gg 1$, we can set $\beta=\frac{7}{4}$ in (\ref{eq:thm0beta}) to minimize the approximation error of G-SAC. We can also set $\beta$ according to (\ref{eq:optbetaExp}) for L-SAC. Note that since $\beta$ in (\ref{eq:optbetaExp}) is a function of $m$, we represent it with an additional subscript $m$, $\beta_m$, in the remainder of this section. %With respect to the definition of relative errors, we also note that if approximation error is minimizes then the 

In Fig.~\ref{FIG:variable2}, we plot average relative error vs. $\lambda$, where $\lambda \in \{10^{-3}, 10^{-2}, 10^{-1}, 1, 10, 10^2, 10^3\}$ and $m=8$. The results for L-SAC when used with Lagrange codes and $\beta_8$ is plotted by the orange solid curve; when $\beta=1$, the result is plotted by the orange dashed curve. Similarly, we plot (green solid curve) the result for G-SAC when $\beta=\frac{7}{4}$ and plot (green dashed curve) the results for $\beta=1$. In both G-SAC methods, we fix $K_1=5$. As shown in Fig.~\ref{FIG:variable2}, when $\lambda \leq 1$, we can lower the average relative error of the G-SAC and L-SAC methods by setting $\beta=1$. In this case, while the $\epsilon$-approximate MatDot codes provide its first estimate, the average relative error of this estimate is lower than those of the G-SAC with $K_1=8$ and L-SAC. A similar effect is also seen in Fig.~\ref{FIG:variable1} for $m=8$, where we used uncorrelated input matrices ($\lambda=0$). On the other hand, Figure~\ref{FIG:variable2} shows that when $\lambda \geq 10$, both G-SAC and L-SAC provide lower average relative errors compared to $\epsilon$-approximate MatDot codes if their $\beta$ are respectively set to $\frac{7}{4}$ and $\beta_8$. 

\section{conclusion}\label{sec:conclusion}
In this paper, we propose group-wise and layer-wise SAC which enable approximate computing in prior CDC schemes. Our SAC schemes extend the approximation procedure into multiple resolution layers beyond the single layer of $\epsilon$-approximate MatDot codes~\cite{9509407}. We analytically and experimentally studied different sources of error in SAC and provided design guidelines. Our simulations justify the superiority of SAC over $\epsilon$-approximate MatDot codes~\cite{9509407} by achieving a better tradeoff between approximation threshold and relative error. Optimally setting parameters in SAC not only provides a lower approximation threshold but also yields a lower relative error when compared to $\epsilon$-approximate MatDot codes. 
%We will study the numerical stability of SAC and apply it to practical computational problems, such as DNN training, in future work.

While our paper focuses on matrix multiplication where polynomial-based CDC  is applied, a direct extension of this paper is to apply SAC to more practical applications such as training deep neural networks. Another extension is to use SAC in a distributed system composed of a heterogeneous set of workers of different computational abilities. We aim to provide optimal parameters of SAC with respect to such a heterogeneous distributed system in future works.  

\bibliographystyle{IEEEtran} 
\bibliography{literature}

% Generated by IEEEtran.bst, version: 1.14 (2015/08/26)
\begin{thebibliography}{10}
\providecommand{\url}[1]{#1}
\csname url@samestyle\endcsname
\providecommand{\newblock}{\relax}
\providecommand{\bibinfo}[2]{#2}
\providecommand{\BIBentrySTDinterwordspacing}{\spaceskip=0pt\relax}
\providecommand{\BIBentryALTinterwordstretchfactor}{4}
\providecommand{\BIBentryALTinterwordspacing}{\spaceskip=\fontdimen2\font plus
\BIBentryALTinterwordstretchfactor\fontdimen3\font minus
  \fontdimen4\font\relax}
\providecommand{\BIBforeignlanguage}[2]{{%
\expandafter\ifx\csname l@#1\endcsname\relax
\typeout{** WARNING: IEEEtran.bst: No hyphenation pattern has been}%
\typeout{** loaded for the language `#1'. Using the pattern for}%
\typeout{** the default language instead.}%
\else
\language=\csname l@#1\endcsname
\fi
#2}}
\providecommand{\BIBdecl}{\relax}
\BIBdecl

\bibitem{lee2017speeding}
K.~Lee, M.~Lam, R.~Pedarsani, D.~Papailiopoulos, and K.~Ramchandran, ``Speeding
  up distributed machine learning using codes,'' \emph{IEEE Trans. Inf.
  Theory}, vol.~64, no.~3, pp. 1514--1529, 2017.

\bibitem{dutta2016short}
S.~Dutta, V.~Cadambe, and P.~Grover, ``Short-dot: Computing large linear
  transforms distributedly using coded short dot products,'' \emph{Int. Conf.
  Neural Inf. Proc. Sys. (NeurIPS)}, vol.~29, 2016.

\bibitem{lee2017high}
K.~Lee, C.~Suh, and K.~Ramchandran, ``High-dimensional coded matrix
  multiplication,'' in \emph{IEEE Int. Symp. Inf. Theory (ISIT)}, 2017, pp.
  2418--2422.

\bibitem{yu2017polynomial}
Q.~Yu, M.~A. Maddah-Ali, and A.~S. Avestimehr, ``Polynomial codes: an optimal
  design for high-dimensional coded matrix multiplication,'' in \emph{Int.
  Conf. Neural Inf. Proc. Sys. (NeurIPS)}, 2017, pp. 4406--4416.

\bibitem{dutta2019optimal}
S.~Dutta, M.~Fahim, F.~Haddadpour, H.~Jeong, V.~Cadambe, and P.~Grover, ``On
  the optimal recovery threshold of coded matrix multiplication,'' \emph{IEEE
  Trans. on Inf. Theory}, vol.~66, no.~1, pp. 278--301, 2019.

\bibitem{kiani2018exploitation}
S.~Kiani, N.~Ferdinand, and S.~C. Draper, ``Exploitation of stragglers in coded
  computation,'' in \emph{IEEE Int. Symp. Inf. Theory (ISIT)}, 2018, pp.
  1988--1992.

\bibitem{kianidehkordi2020hierarchical}
------, ``Hierarchical coded matrix multiplication,'' \emph{IEEE Trans. Inf.
  Theory}, vol.~67, no.~2, pp. 726--754, 2020.

\bibitem{mallick2019rateless}
A.~Mallick, M.~Chaudhari, U.~Sheth, G.~Palanikumar, and G.~Joshi, ``Rateless
  codes for near-perfect load balancing in distributed matrix-vector
  multiplication,'' \emph{Proc. of the ACM on Measurement and Analysis of
  Computing Sys}, vol.~3, no.~3, pp. 1--40, 2019.

\bibitem{yu2020straggler}
Q.~Yu, M.~A. Maddah-Ali, and A.~S. Avestimehr, ``Straggler mitigation in
  distributed matrix multiplication: Fundamental limits and optimal coding,''
  \emph{IEEE Trans. Inf. Theory}, vol.~66, no.~3, pp. 1920--1933, 2020.

\bibitem{dutta2018unified}
S.~Dutta, Z.~Bai, H.~Jeong, T.~M. Low, and P.~Grover, ``A unified coded deep
  neural network training strategy based on generalized polydot codes,'' in
  \emph{IEEE Int. Symp. Inf. Theory (ISIT)}.\hskip 1em plus 0.5em minus
  0.4em\relax IEEE, 2018, pp. 1585--1589.

\bibitem{yu2019lagrange}
Q.~Yu, S.~Li, N.~Raviv, S.~M.~M. Kalan, M.~Soltanolkotabi, and S.~A.
  Avestimehr, ``Lagrange coded computing: Optimal design for resiliency,
  security, and privacy,'' in \emph{Int. Conf. on Artificial Intelligence and
  Statistics (AISTATS)}, 2019, pp. 1215--1225.

\bibitem{quarteroni2010numerical}
A.~Quarteroni, R.~Sacco, and F.~Saleri, \emph{Numerical mathematics}.\hskip 1em
  plus 0.5em minus 0.4em\relax Springer Science \& Business Media, 2010,
  vol.~37.

\bibitem{fahim2021numerically}
M.~Fahim and V.~R. Cadambe, ``Numerically stable polynomially coded
  computing,'' \emph{IEEE Trans. on Inf. Theory}, vol.~67, no.~5, pp.
  2758--2785, 2021.

\bibitem{charalambides2021approximate}
N.~Charalambides, M.~Pilanci, and A.~O. Hero, ``Approximate weighted cr coded
  matrix multiplication,'' in \emph{IEEE Int. Conf. Acoustics, Speech and
  Signal Processing (ICASSP)}, 2021, pp. 5095--5099.

\bibitem{chang2019random}
W.-T. Chang and R.~Tandon, ``Random sampling for distributed coded matrix
  multiplication,'' in \emph{IEEE Int. Conf. Acoustics, Speech and Signal
  Processing (ICASSP)}, 2019, pp. 8187--8191.

\bibitem{ferdinand2016anytime}
N.~S. Ferdinand and S.~C. Draper, ``Anytime coding for distributed
  computation,'' in \emph{IEEE Annual Allerton Conf. on Commun., Control, and
  Comput.}, 2016, pp. 954--960.

\bibitem{zhu2017sequential}
J.~Zhu, Y.~Pu, V.~Gupta, C.~Tomlin, and K.~Ramchandran, ``A sequential
  approximation framework for coded distributed optimization,'' in \emph{IEEE
  Annual Allerton Conf. on Commun., Control, and Comput.}, 2017, pp.
  1240--1247.

\bibitem{jahani2021codedsketch}
T.~Jahani-Nezhad and M.~A. Maddah-Ali, ``Codedsketch: A coding scheme for
  distributed computation of approximated matrix multiplication,'' \emph{IEEE
  Trans. Inf. Theory}, vol.~67, no.~6, pp. 4185--4196, 2021.

\bibitem{gupta2018oversketch}
V.~Gupta, S.~Wang, T.~Courtade, and K.~Ramchandran, ``Oversketch: Approximate
  matrix multiplication for the cloud,'' in \emph{IEEE Int. Conf. on Big Data},
  2018, pp. 298--304.

\bibitem{9509407}
H.~Jeong, A.~Devulapalli, V.~R. Cadambe, and F.~P. Calmon,
  ``$\epsilon-$approximate coded matrix multiplication is nearly twice as
  efficient as exact multiplication,'' \emph{IEEE J. Sel. Areas Inf. Theory},
  vol.~2, no.~3, pp. 845--854, 2021.

\bibitem{drineas2006fast}
P.~Drineas, R.~Kannan, and M.~W. Mahoney, ``Fast {Monte} {Carlo} algorithms for
  matrices {I}: Approximating matrix multiplication,'' \emph{SIAM J. on
  Comput.}, vol.~36, no.~1, pp. 132--157, 2006.

\bibitem{ramamoorthy2021numerically}
A.~Ramamoorthy and L.~Tang, ``Numerically stable coded matrix computations via
  circulant and rotation matrix embeddings,'' in \emph{IEEE Int. Symp. Inf.
  Theory (ISIT)}, 2021, pp. 1712--1717.

\end{thebibliography}

\newpage

\appendix
\section{Appendix}
\subsection{Admissible Ranges for Number of Resolution Layers} \label{app:Lrange}
In this section, we prove the maximum and minimum admissible bounds on the number of resolution layers in group-wise SAC.

\begin{claim}
In group-wise SAC, the number of layers is in the ranges $R_{\text{G-SAC}}-K \leq L_{\text{S$\epsilon$AMD}} \leq R_{\text{G-SAC}}-1$.
\end{claim}
 
\textit{Proof:} Since the exact recovery is achieved at $R_{\text{G-SAC}}$ and the approximation threshold increases as $R_{\text{G-SAC},l}=R_{\text{G-SAC},l-1}+1$ for each $l \in \{2,3,\ldots L_{\text{G-SAC}}\}$, the following formula holds for the last resolution layer. We have $R_{\text{G-SAC},L_{\text{G-SAC}}}=R_{\text{G-SAC},1}+L_{\text{G-SAC}}-1$ which equals to $R_{\text{G-SAC}}-1$. Therefore, $L_{\text{G-SAC}}=R_{\text{G-SAC}}-R_{\text{G-SAC},1}$. Since $R_{\text{G-SAC},1} \in [K]$, we can conclude that $L_{\text{G-SAC}} \in \{R_{\text{G-SAC}}-K, \ldots, R_{\text{G-SAC}}-1\} $. $\blacksquare$

\subsection{Proof of Thm.~\ref{thm0}}\label{app:thm0}
Recalling the notation from Sec.~\ref{sec:SAC}, in resolution layer $l \in [L_{\text{G-SAC}}]$ of group-wise SAC, the master recovers ${C}_l=\sum_{k=1}^{m_l} A_{i_k}B_{i_k}$ to some accuracy. Also, assume that the $A_{i_1}B_{i_1},\ldots,A_{i_K}B_{i_K}$ products are a {\em uniform} random permutation of $A_1B_1,\ldots,A_{K}B_{K}$. We first calculate the expected value of ${C}_l$, proving that the scaled sum $\frac{K}{m_l}{C}_l$ is an unbiased estimate of $C=AB$. We compute $\mathbb{E}\left(\frac{K}{m_l}\sum_{k=1}^{m_l} A_{I_k}B_{I_k}\right) $, where the expectation is taken with respect to the random indices $\{I_k\}_{k=1}^K$. 
\begin{align}\label{app:eq:thm1}
&\mathbb{E}\left(\frac{K}{m_l}\sum_{k=1}^{m_l} A_{I_k}B_{I_k}\right)  \nonumber \\ &\stackrel{\mathclap{\scriptsize	\mbox{(1)}}}{=} \frac{K}{m_l}\sum_{i_1,\ldots,i_{K}} \left(\left( \sum_{k=1}^{m_l} A_{i_k}B_{i_k}\right) P(I_1=i_1,\ldots,I_K = i_{K})\right) \nonumber \\ &\stackrel{\mathclap{\scriptsize	\mbox{(2)}}}{=} \frac{K}{m_l} \sum_{j=1}^K A_jB_j P(j\in \{I_k\}_{k=1}^{m_l}) \nonumber \\ &\stackrel{\mathclap{\scriptsize	\mbox{(3)}}}{=}  \frac{K}{m_l} \sum_{j=1}^K A_jB_j \left(1 - P\left(j\in \{I_k\}_{k=m_l+1}^K\right) \right) \nonumber \\ &\stackrel{\mathclap{\scriptsize	\mbox{(4)}}}{=} \frac{K}{m_l} \sum_{j=1}^K A_jB_j \left(1-\frac{{K-1 \choose m_l}}{{K \choose m_l}} \right) \nonumber \\ &\stackrel{\mathclap{\scriptsize	\mbox{(5)}}}{=} \frac{K}{m_l}\sum_{j=1}^K A_jB_j \frac{m_l}{K} = \sum_{j=1}^K A_jB_j = AB. 
\end{align}
%\begin{align}\label{app:eq:thm1}
%\mathbb{E}\left(\frac{K}{m_l}\sum_{k=1}^{m_l} A_{I_k}B_{I_k}\right)  \\ &\stackrel{\mathclap{\scriptsize	\mbox{(1)}}}{=} \frac{K}{m}\sum_{i_1,\ldots,i_{K}} \left(\left( \sum_{k=1}^{m} A_{i_k}B_{i_k}\right) P(i_1,\ldots,i_{K})\right) \nonumber \\ &\stackrel{\mathclap{\scriptsize	\mbox{(2)}}}{=} \frac{K}{m} \sum_{k=1}^K A_kB_k P(k\in \{i_k\}_{k=1}^m) \nonumber \\ &\stackrel{\mathclap{\scriptsize	\mbox{(3)}}}{=}  \frac{K}{m} \sum_{k=1}^K A_kB_k \left(P\left(k\in \{i_k\}_{k=1}^K\right) - P\left(k\in \{i_k\}_{k=m+1}^K\right) \right) \nonumber \\ &\stackrel{\mathclap{\scriptsize	\mbox{(4)}}}{=} \frac{K}{m} \sum_{k=1}^K A_kB_k \left(\frac{{K \choose m} - {K-1 \choose m}}{{K \choose m}} \right) \nonumber \\ &\stackrel{\mathclap{\scriptsize	\mbox{(5)}}}{=} \frac{K}{m}\sum_{k=1}^K A_kB_k \frac{m}{K} = \sum_{k=1}^K A_kB_k = AB. 
%\end{align}
% two col
In~(\ref{app:eq:thm1}), the first equality is obtained by expanding the expected value as the weighted sum of the $\frac{K}{m_l}\sum_{k=1}^{m_l} A_{i_k}B_{i_k}$ values, with the probabilities $P(I_1=i_1,\ldots,I_K = i_{K})$ as the weights. In this expansion, the sample $(i_1,\ldots,i_k)$ is the realization of $K$ random variables $(I_1,\ldots,I_K)$, whose values depend on the indices of the random permutation of $A_1B_1,\ldots,A_{K}B_{K}$. The second equality of~(\ref{app:eq:thm1}) is a result of rewriting $\sum_{i_1,\ldots,i_K, j\in \{i_k\}_{k=1}^{m_l}} P(I_1=i_1,\ldots,I_K=i_K)$ by $P(j \in \{I_k\}_{k=1}^{m_l})$. In the third equality of~(\ref{app:eq:thm1}), the probability of the event $j \in \{I_k\}_{k=1}^{m_l}$ occurs is replaced with 1 minus the probability of the event does not occur. In the forth equality of~(\ref{app:eq:thm1}), ${K \choose m_l}$ is the total number of ways that $m_l$ (distinct) indices are picked first from $\{1,\ldots, K\}$. If index $j$ is excluded, then there are $K-1$ other indices and ${K-1 \choose m_l}$ ways to choose the first $m_l$ indices from $\{1,\ldots, K\}/\{j\}$. In the last equations, we simplify the expressions and prove that $\mathbb{E}\left(\frac{K}{m_l}\sum_{k=1}^{m_l} A_{I_k}B_{I_k}\right)$ equals $AB$ in~(\ref{app:eq:thm1}).

We next solve for the scaling $\beta$ that is the solution to $\operatorname*{argmin}_\beta \mathbb{E}\left(\left\| C-\beta {C}_l\right\|_F^2 \right)$. Since the problem is convex is $\beta$, we set the derivative to zero and find $\beta^*$. We start by expanding $\mathbb{E}\left(\| C-\beta C_l\|_F^2 \right)$ as 
\begin{align}\label{app:eq2:thm1}
&\| C\|_F^2 - 2\beta \Tr\left(C^T\mathbb{E}\left(C_l\right)\right) + \beta^2 \mathbb{E}\left(\| C_l\|_F^2\right) \nonumber\\ &\stackrel{\mathclap{\scriptsize	\mbox{(1)}}}{=} \| C\|_F^2 \left(1 - 2\beta \frac{m_l}{K} \right)  +  \beta^2 \mathbb{E}\left(\| C_l\|_F^2\right) \nonumber\\ &\stackrel{\mathclap{\scriptsize	\mbox{(2)}}}{=} \| C\|_F^2 \left(1 - 2\beta \frac{m_l}{K} \right)  + \beta^2 \left(\sum_{j=1}^K \left( \| A_jB_j\|_F^2 P(j\in \{i_k\}_{k=1}^{m_l})\right) \right. \nonumber\\ &+ \left. 2 \sum_{\substack{j,j'=1 \\ j' < j}}^K \left(\Tr\left(\left(A_{j'}B_{j'}\right)^T\left(A_jB_j\right)\right)P(j,j'\in \{i_k\}_{k=1}^{m_l})\right)   \right) \nonumber\\ &\stackrel{\mathclap{\scriptsize	\mbox{(3)}}}{=} \| C\|_F^2 \left(1 - 2\beta \frac{m_l}{K} \right)  + \beta^2 \left(\sum_{j=1}^K \left( \| A_jB_j\|_F^2 \frac{{K \choose m_l} - {K-1 \choose m_l}}{{K \choose m_l}}\right) \right. \nonumber \\ &+ \left. 2 \sum_{\substack{j,j'=1 \\ j' < j}}^K \left(\Tr\left(\left(A_{j'}B_{j'}\right)^T\left(A_jB_j\right)\right)\frac{{K \choose m_l} - 2 {K-1 \choose m_l} + {K-2 \choose m_l}}{{K \choose m_l}}\right) \right) \nonumber\\ &\stackrel{\mathclap{\scriptsize	\mbox{(4)}}}{=} \| C\|_F^2 \left(1 - 2\beta \frac{m_l}{K} \right)  + \beta^2 \left(M_1 \frac{m_l}{K}+ 2 M_2\frac{m_l(m_l-1)}{K(K-1)}\right)   ,
\end{align}
%\begin{align*}
%&\| C\|_F^2 - 2\beta \Tr\left(C^T\mathbb{E}\left(C_m\right)\right) + \beta^2 \mathbb{E}\left(\| C_m\|_F^2\right) \\ &= \| C\|_F^2 \left(1 - 2\beta \frac{m}{K} \right)  +  \beta^2 \mathbb{E}\left(\| C_m\|_F^2\right)\\ &= \| C\|_F^2 \left(1 - 2\beta \frac{m}{K} \right)  + \beta^2 \left(\sum_{k=1}^K \left( \| A_kB_k\|_F^2 P(k\in \{i_k\}_{k=1}^m)\right) \right. \\ &+ \left. 2 \sum_{\substack{i,j=1 \\ i < j}}^K \left(\Tr\left(\left(A_iB_i\right)^T\left(A_jB_j\right)\right)P(i,j\in \{i_k\}_{k=1}^m)\right)   \right) \\ &= \| C\|_F^2 \left(1 - 2\beta \frac{m}{K} \right)  + \beta^2 \left(\sum_{k=1}^K \left( \| A_kB_k\|_F^2 \frac{{K \choose m} - {K-1 \choose m}}{{K \choose m}}\right) \right. \\ &+ \left. 2 \sum_{\substack{i,j=1 \\ i < j}}^K \left(\Tr\left(\left(A_iB_i\right)^T\left(A_jB_j\right)\right)\frac{{K \choose m} - 2 {K-1 \choose m} + {K-2 \choose m}}{{K \choose m}}\right) \right) \\ &= \| C\|_F^2 \left(1 - 2\beta \frac{m}{K} \right)  + \beta^2 \left(M_1 \frac{m}{K}+ 2 M_2\frac{m(m-1)}{K(K-1)}\right)   ,
%\end{align*}
% two col
where
\begin{align*}
M_1 &= \sum_{j=1}^{K} \| A_jB_j\|_F^2, \text{and} \\
M_2 &= \sum_{j,j'=1, j' < j}^K \Tr\left( \left(A_{j'}B_{j'}\right)^T \left(A_jB_j \right)\right).
\end{align*}
In the first equation of~(\ref{app:eq2:thm1}), we use~(\ref{app:eq:thm1}) to replace $\mathbb{E}(C_l)$ with $\frac{m_l}{K}C$. The second equation is obtained by expanding $\mathbb{E}(\| C_l \|_F^2)$. To do this expansion, we use the results of second equation in~(\ref{app:eq:thm1}). Similar to the logic we used in the third and forth equations of~(\ref{app:eq:thm1}), in the third equation of~(\ref{app:eq2:thm1}) we replace $P(j\in \{i_k\}_{k=1}^{m_l})$ with $\frac{{K \choose m_l} - {K-1 \choose m_l}}{{K \choose m_l}}$. We also replace $P(j,j'\in \{i_k\}_{k=1}^{m_l})$ with $\frac{{K \choose m_l} - 2 {K-1 \choose m_l} + {K-2 \choose m_l}}{{K \choose m_l}}$. The intuition behind the latter replacement is similar to the former. The probability of the event $j,j' \in \{I_k\}_{k=1}^{m_l}$ occurs is equal to 1 minus the probability of at least one of $j$ or $j'$ are excluded plus the probability of both indices are excluded. If $j$ (or $j'$) are excluded, we showed that there are ${K-1 \choose m_l}$ ways to choose the first $m_l$ indices from $[K]/\{j\}$ (or $[K]/\{j'\}$). If both indices $j$ and $j'$ are excluded, then there are $K-2$ other indices and ${K-2 \choose m_l}$ ways to choose the first $m_l$ indices from $[K]/\{j,j'\}$. Combining these results together, we conclude $P(j,j'\in \{i_k\}_{k=1}^{m_l})=\frac{{K \choose m_l} - 2 {K-1 \choose m_l} + {K-2 \choose m_l}}{{K \choose m_l}}$. Finally, with simplifying the mathematical expressions we obtain the fourth equation in~(\ref{app:eq2:thm1}).  

Next, we take the derivative of $\mathbb{E}\left(\| C-\beta C_l\|_F^2 \right)$ with respect to $\beta$ and set the result to zero. Thus, we have
\begin{align*}
&-2\| C\|_F^2\frac{m_l}{K} + 2\beta^* \left(M_1 \frac{m_l}{K}+ 2 M_2\frac{m_l(m_l-1)}{K(K-1)}\right) =0 \\ & \Rightarrow \beta^*= \frac{\| C\|_F^2\frac{m_l}{K}}{\left(M_1 \frac{m_l}{K}+ 2 M_2\frac{m_l(m_l-1)}{K(K-1)}\right)}.
\end{align*}
%\begin{align*}
%&-2\| C\|_F^2\frac{m}{K} + 2\beta^* \left(M_1 \frac{m}{K}+ 2 M_2\frac{m(m-1)}{K(K-1)}\right) =0\\ &\Rightarrow \beta^*= \frac{\| C\|_F^2\frac{m}{K}}{\left(M_1 \frac{m}{K}+ 2 M_2\frac{m(m-1)}{K(K-1)}\right)}.
%\end{align*}
% two col
Since $\| C\|_F^2=M_1+2M_2$, we have
\begin{align*}
\beta^*= \frac{M_1\frac{m_l}{K} + 2M_2\frac{m_l}{K}}{\left(M_1 \frac{m_l}{K}+ 2 M_2\frac{m_l(m_l-1)}{K(K-1)}\right)} = \frac{M_1 + 2M_2}{\left(M_1 + 2 M_2\frac{(m_l-1)}{(K-1)}\right)}. \; \blacksquare
\end{align*}

%\subsubsection{Proof to Theorem~\ref{thm1}}
%To prove Thm.~\ref{thm1}, we first prove the following lemmas.
%\begin{lem}\label{lem1}
%If $l$ balls are drawn at random without replacement from a box containing $n$ blue and $N-n$ red balls, then the expected number of blue drawn balls is equal to $\frac{nl}{N}$.
%\end{lem}
%\textit{Proof:} Let $m$ be a random variable counting the number of blue balls out of $l$ drawn balls. The expected value of $m$ is 
%\begin{align}\label{eq1:lem1}
%\mathbb{E}(m) = \sum_{i=0}^l i P(m=i) = \sum_{i=0}^l \frac{i{n \choose i} {N-n \choose l-i}}{{N \choose l}}
%\end{align}
%One can easily show that the numerator of each term in above summation can be rewritten as $i{n \choose i} {N-n \choose l-i} = n{n-1 \choose i-1 }{N-n \choose l-i}$. Therefore, using Vandermonde's identity~\cite{paolillo2017application}, the right-hand-side of equation (\ref{eq1:lem1}) can be simplified  into 
%\begin{align*}
%\frac{n}{{N \choose l}} \sum_{j=0}^{l-1} {n-1 \choose j}{N-n \choose l-j-1} = \frac{n}{{N \choose l}} {N-1 \choose l-1} = \frac{nl}{N} \;\; \blacksquare 
%\end{align*}

\subsection{Proof of Thm.~\ref{thm1}}\label{app:thm1}
 Recalling from Sec.~\ref{SEC:rsac}, in layer-wise SAC the master recovers ${C}_m = \sum_{j=1}^{K} \alpha_{j}\tilde{S}_A(y_{j})\tilde{S}_B(y_{j})\mathbbm{1}_{m_j > 0}$ to some accuracy when $m$ workers report in. Recall $m=\sum_{j=1}^K m_j$, where $m_j$ is the number of workers that complete the evaluation of $\tilde{S}_A(y_{j})\tilde{S}_B(y_{j})$ to some accuracy. We next solve $\operatorname*{argmin}_\beta \mathbb{E}\left(\| C-\beta {C}_m\|_F^2 \right)$, where the expectation is taken with respect to the random $m_j$. To do this, we first calculate $\mathbb{E}({C}_m)$. 
\begin{align*}
\mathbb{E}({C}_m) &= \sum_{j=1}^K \alpha_j \tilde{S}_A(y_j)\tilde{S}_B(y_j) \mathbb{E}(\mathbbm{1}_{m_j > 0}) \\ &= \sum_{j=1}^K \alpha_j \tilde{S}_A(y_j)\tilde{S}_B(y_j)\gamma_j
\end{align*}
%\begin{align*}
%\mathbb{E}({C}_m) &= \sum_{k=1}^K \alpha_k \tilde{S}_A(y_k)\tilde{S}_B(y_k) \mathbb{E}(\mathbbm{1}_{m_k > 0}) \\&= \sum_{k=1}^K \alpha_k \tilde{S}_A(y_k)\tilde{S}_B(y_k)\gamma_k
%\end{align*}
% two col
where $\gamma_j = P(m_j > 0)=1 - {N-n_j  \choose m} / {N \choose m}$. This equality holds with respect to the logic we used to prove the forth equation in~(\ref{app:eq:thm1}). Using the definition of $C_m$, we can expand $\mathbb{E}\left(\| C-\beta {C}_m\|_F^2\right)$ as
\begin{align}\label{thm2eq2}
&\| C\|_F^2 - 2\beta \Tr\left(C^T\mathbb{E}\left({C}_m\right)\right) + \beta^2 \mathbb{E}\left(\|{C}_m\|_F^2\right) = \| C\|_F^2  + \nonumber \\  &-2\beta \sum_{j=1}^K \left[\Tr\left(\left( \tilde{S}_A(y_j)\tilde{S}_B(y_j)\right)^T\left( \tilde{S}_A(y_j)\tilde{S}_B(y_j)\right)\right)\alpha_j^2\gamma_j \right]\nonumber  \\ &-2\beta \sum_{\substack{j,j'=1 \\ j' < j}}^K  \left[\Tr\left(\left( \tilde{S}_A(y_{j'})\tilde{S}_B(y_{j'})\right)^T\left( \tilde{S}_A(y_j)\tilde{S}_B(y_j)\right)\right) \right. \nonumber \\ &\times \left. \alpha_{j'}\alpha_j\gamma_{{j'}}\right] \nonumber \\ &-2\beta \sum_{\substack{j,j'=1 \\ j' < j}}^K  \left[\Tr\left(\left(\tilde{S}_A(y_{j'})\tilde{S}_B(y_{j'})\right)^T\left( \tilde{S}_A(y_j)\tilde{S}_B(y_j)\right)\right) \nonumber \right. \\ &\times \left. \alpha_{j'}\alpha_j \gamma_{j}\right] \nonumber \\ &+  \beta^2 \sum_{j=1}^K \left[\Tr\left(\left( \tilde{S}_A(y_j)\tilde{S}_B(y_j)\right)^T\left(\tilde{S}_A(y_j)\tilde{S}_B(y_j)\right)\right)\alpha_j^2\gamma_j\right] \nonumber\\ &+ 2\beta^2 \sum_{\substack{j,j'=1 \\ j' < j}}^K \left[ \Tr\left(\left( \tilde{S}_A(y_{j'})\tilde{S}_B(y_{j'})\right)^T\left( \tilde{S}_A(y_j)\tilde{S}_B(y_j)\right)\right) \nonumber \right. \\ &\times \left. \alpha_{j'}\alpha_j\gamma_{j',j}\right],
\end{align}
where $\gamma_{j',j} = \mathbb{E}(\mathbbm{1}_{m_{j'},m_j > 0})=P(m_{j'},m_j>0)$. Using similar logic to that of third equation in~(\ref{app:eq2:thm1}), we can conclude that  
\begin{align*}
\gamma_{j',j}  = \frac{{N \choose m} - {N-n_{j'} \choose m} - {N-n_j \choose m} + {N-n_{j'}-n_j \choose m}}{{N \choose m}} .
\end{align*}
Letting $\tilde{M}_{i,j} = \alpha_i\alpha_j \Tr\left( \left( \tilde{S}_A(y_i) \tilde{S}_B(y_i)\right)^T \left(\tilde{S}_A(y_j)\tilde{S}_B(y_j)\right)\right)$ and $\tilde{M}_i = \alpha_i^2 \|\tilde{S}_A(y_i)\tilde{S}_B(y_i)\|_F^2$, we can rewrite~(\ref{thm2eq2}) as
\begin{align*}
&\mathbb{E}\left(\| C-\beta {C}_m\|_F^2\right) \\ &= \| C\|_F^2 - 2\beta \sum_{j=1}^K \tilde{M}_{j}\gamma_j  - 2\beta \sum_{\substack{j,j'=1 \\ j' < j}}^K \tilde{M}_{j',j} (\gamma_{j'} + \gamma_j)\\&+ \beta^2 \sum_{j=1}^K \tilde{M}_{j}\gamma_j + 2\beta^2 \sum_{\substack{j,j'=1 \\ j' < j}}^K \tilde{M}_{j',j}\gamma_{j',j}.
\end{align*}
%\begin{align*}
%&\mathbb{E}\left(\| C-\beta {C}_m\|_F^2\right) = \| C\|_F^2 - 2\beta \sum_{k=1}^K \tilde{M}_{k}\gamma_k + \\ & - 2\beta \sum_{\substack{i,j=1 \\ i < j}}^K \tilde{M}_{i,j} (\gamma_i + \gamma_j)+ \beta^2 \sum_{k=1}^K \tilde{M}_{k}\gamma_k + 2\beta^2 \sum_{\substack{i,j=1 \\ i < j}}^K \tilde{M}_{i,j}\gamma_{i,j}.
%\end{align*}
% two col
To solve $\operatorname*{argmin}_\beta \mathbb{E}\left(\| C-\beta {C}_m\|_F^2 \right)$, we take the derivative of $\mathbb{E}\left(\| C-\beta {C}_m\|_F^2\right)$ with respect to $\beta$ and set it to 0, optimizing $\beta$ via
\begin{align*}
&- 2 \sum_{j=1}^K \tilde{M}_{j}\gamma_j - 2 \sum_{\substack{j,j'=1 \\ j' < j}}^K \tilde{M}_{j,j'} (\gamma_{j'} + \gamma_j) + 2\beta^* \sum_{j=1}^K \tilde{M}_{j}\gamma_j + \\ &+ 4\beta^* \sum_{\substack{j,j'=1 \\ j' < j}}^K \tilde{M}_{j',j}\gamma_{j',j} = 0 \\
&\Rightarrow \beta^* = \frac{ \sum_{j=1}^K \tilde{M}_{j}\gamma_j + \sum_{\substack{j,j'=1 \\ j' < j}}^K \tilde{M}_{j',j} (\gamma_{j'}+\gamma_j)}{ \sum_{j=1}^K \tilde{M}_{j}\gamma_j +  2\sum_{\substack{j,j'=1 \\ j' < j}}^K \tilde{M}_{j',j}\gamma_{j',j}}. \; \blacksquare
\end{align*}

\subsection{Proof to $\lim_{\epsilon \to 0} \tilde{C}_m = {C}_m$ in layer-wise SAC}\label{app:thm:eq}
Recall from Sec.~\ref{SEC:rsac} that in layer-wise SAC the master can compute 
\begin{align*}
\tilde{C}_m= \sum_{k=1}^{K} \alpha_{k} \frac{\sum_{i=1}^{m_k}\tilde{S}_A(z_{k,j_{k,i}})\tilde{S}_B(z_{k,j_{k,i}})}{m_{k}}.
\end{align*}
when the $m$ fastest workers ($m=\sum_{k=1}^K m_k$) finish computing the $\tilde{S}_A(z_{k,j_{k,i}})\tilde{S}_B(z_{k,j_{k,i}})$ products, $k\in [K]$ and $i \in [m_k]$. Also, recall that $z_{k,i}$ is $\epsilon$-close to $y_k$, for any $k\in [K]$ and all $i\in [n_k]$. We use the following lemma to prove that $\lim_{\epsilon \to 0} \tilde{C}_m = {C}_m$, where ${C}_m = \sum_{k=1}^{K} \alpha_{k}\tilde{S}_A(y_{k})\tilde{S}_B(y_{k})\mathbbm{1}_{m_k > 0}$. 

\begin{lem} \label{lem2}
Let $y$ be $\epsilon$-close to $x$ and assume that the $n-$degree polynomial $f(x)=c_0 + c_1 x + \ldots + c_n x^n$ is evaluated at $x$. Both $x$ and the coefficients $c_i$ are bounded as $|x| \leq \lambda_1$ and $|c_i|\leq \lambda_2$. The error of approximating $f(y)$ with $f(x)$ can be bounded as
\begin{align*}
|f(y)-f(x)| \leq \lambda_2 \left(\frac{(\lambda_1+\epsilon)^{n+1}-(\lambda_1+\epsilon)}{(\lambda_1+\epsilon)-1} - \frac{\lambda_1^{n+1}-\lambda_1}{\lambda_1-1}\right) . 
\end{align*}  
\end{lem}

\textit{Proof of Lem.~\ref{lem2}:} Using the definition of $f(x)$ and $f(y)$, we first expand the error as 
\begin{align}\label{app:lem1:eq1}
\left|f(y)-f(x)\right| = \left|\sum_{i=1}^{n} c_i (y-x)^i\right|
\end{align} 
% \begin{align*}
%&\left|f(y)-f(x)\right| = \left|\sum_{i=1}^{n} c_i (y-x)^i\right| \leq \lambda_2 \epsilon \sum_{i=1}^n \sum_{j=0}^{i-1} |y|^j|x|^{i-1-j} \\ &\leq \lambda_2 \epsilon \sum_{i=1}^n \sum_{j=0}^{i-1} (\lambda_1+\epsilon)^j\lambda_1^{i-1-j} = \lambda_2 \epsilon \sum_{i=1}^n \lambda_1^{i-1} \frac{(1+\frac{\epsilon}{\lambda_1})^i-1}{\frac{\epsilon}{\lambda_1}} \\ &= \lambda_2 \left(\frac{(\lambda_1+\epsilon)^{n+1}-(\lambda_1+\epsilon)}{(\lambda_1+\epsilon)-1} - \frac{\lambda_1^{n+1}-\lambda_1}{\lambda_1-1}\right). \;\; \blacksquare
%\end{align*} 
% two col
Replacing $(y-x)^i$ with its expansion $ (y-x)\left(\sum_{j=0}^{i-1} y^jx^{i-1-j}\right)$ and using triangular inequality, we can bound~(\ref{app:lem1:eq1}) as
\begin{align}\label{app:lem1:ineq1}
\left|f(y)-f(x)\right| \leq \lambda_2 \epsilon \sum_{i=1}^n \sum_{j=0}^{i-1} |y|^j|x|^{i-1-j}.
\end{align}
Note that in the above inequality, we use the bounds for $|c_i|\leq \lambda_2$ and $|y-x| \leq \epsilon$. Using the bounds for $|x|\leq \lambda_1$ and $|y| \leq |x|+\epsilon \leq \lambda_1+\epsilon$, we can further upper bound the right-hand-side of~(\ref{app:lem1:ineq1}). In the following we provide this upper bound.  We then in the first and second equations simplify the mathematical expressions using the close form of geometric series,   and prove the lemma.
\begin{align*}
&\left|f(y)-f(x)\right| \leq \lambda_2 \epsilon \sum_{i=1}^n \sum_{j=0}^{i-1} (\lambda_1+\epsilon)^j\lambda_1^{i-1-j} \\ &\stackrel{\mathclap{\scriptsize	\mbox{(1)}}}{=} \lambda_2 \epsilon \sum_{i=1}^n \lambda_1^{i-1} \frac{(1+\frac{\epsilon}{\lambda_1})^i-1}{\frac{\epsilon}{\lambda_1}} \\ &\stackrel{\mathclap{\scriptsize	\mbox{(2)}}}{=} \lambda_2 \left(\frac{(\lambda_1+\epsilon)^{n+1}-(\lambda_1+\epsilon)}{(\lambda_1+\epsilon)-1} - \frac{\lambda_1^{n+1}-\lambda_1}{\lambda_1-1}\right). \;\; \blacksquare
\end{align*}

Let $A_{i,j}$ denotes the $(i,j)$th element of matrix $A$ and $B_{i',j'}$ denotes the $(i',j')$th element of matrix $B$, where $i \in [N_x], j' \in [N_y]$, and $j,i' \in [N_z]$. We now set the maximum value of all $|A_{i,j}|$ (and $|B_{i',j'}|$) as an upper bound to the absolute value of each entry in the $A_k$ and $B_k$ submatrices. We bound each $y_k$ by their largest absolute value, $\max_k |y_k|$. These settings let us apply Lem.~\ref{lem2} to the decoding polynomial, $\tilde{S}_A(x)\tilde{S}_B(x)$. Using the same notation as Lem.~\ref{lem2}, we use $\lambda_1$ to denote $\max_k |y_k|$ and $\lambda_2$ to denote $(\max_{i,j} |A_{i,j}|)(\max_{i',j'} |B_{i',j'}|)$. Since the $m$ fastest workers evaluate the $\tilde{S}_A(x)\tilde{S}_B(x)$ polynomial at $x\in \{z_{k,j_{k,i}}\}_{k \in [K], i \in [m_k]}$, we bound $|\tilde{C}_m - {C}_m|$ from above as follows. We first use the triangle inequality to upper bound $|\tilde{C}_m - {C}_m|$ as
\begin{align}\label{up1}
\sum_{k=1}^{K} |\alpha_{k}| \frac{\sum_{i=1}^{m_k}|\tilde{S}_A(z_{k,j_{k,i}})\tilde{S}_B(z_{k,j_{k,i}})-\tilde{S}_A(y_k)\tilde{S}_B(y_k)|}{m_{k}}\mathbbm{1}_{m_k > 0}.
\end{align}
Note that in above inequality, we use the definitions $\tilde{C}_m= \sum_{k=1}^{K} \alpha_{k} \frac{\sum_{i=1}^{m_k}\tilde{S}_A(z_{k,j_{k,i}})\tilde{S}_B(z_{k,j_{k,i}})}{m_{k}}$ and ${C}_m= \sum_{k=1}^{K} \alpha_{k} \frac{\sum_{i=1}^{m_k}\tilde{S}_A(y_k)\tilde{S}_B(y_k)}{m_{k}}$ from Sec.~\ref{SEC:rsac}.
Using Lem.~\ref{lem2}, we further upper bound (\ref{up1}) as
\begin{align*}
&\sum_{k=1}^{K}  \frac{|\alpha_{k}|}{m_{k}}\lambda_2 \left(\frac{(\lambda_1+\epsilon)^{2K-2}-(\lambda_1+\epsilon)}{(\lambda_1+\epsilon)-1} - \frac{\lambda_1^{2K-2}-\lambda_1}{\lambda_1-1}\right)\mathbbm{1}_{m_k > 0}\\
&\leq \lambda_2 \left(\frac{(\lambda_1+\epsilon)^{2K-2}-(\lambda_1+\epsilon)}{(\lambda_1+\epsilon)-1} - \frac{\lambda_1^{2K-2}-\lambda_1}{\lambda_1-1}\right) \sum_{k=1}^{K}  |\alpha_{k}|.
\end{align*}
As $\epsilon$ goes to zero, the right-hand-side of the above inequality converges to 0, thus $\lim_{\epsilon \to 0} \tilde{C}_m = {C}_m$. $\blacksquare$

\subsection{Recovery threshold of group-wise SAC} \label{app:RT}
In this section, we first prove that when multi-group SAC uses only $D=2$ groups, then its recovery threshold is equal to that of two-group SAC. We then prove that if $D>2$,  $R_{\text{G-SAC}}>2K-1$. Substituting $D=2$ in $R_{\text{G-SAC}}=(\sum_{d=1}^D 2^{D-d} K_d)+K_D-1$, we get to $R_{\text{G-SAC}}=(2K_1+K_2)+K_2-1=2K-1$ which equals to that of two-group SAC. When $D>2$, 
\begin{align*}
R_{\text{G-SAC}}&=(\sum_{d=1}^D 2^{D-d} K_d)+K_D-1 \\ &\geq \left(K_D + \sum_{d=1}^{D-1} 2 K_d\right) +K_D-1 \\ &= 2\left(\sum_{d=1}^D K_d \right)-1 =2K-1. \; \blacksquare
\end{align*}

\vfill

%\includepdf[pages={34-40},pagecommand={}]{main.pdf}

\end{document}